\documentclass[10pt, letterpaper, journal]{IEEEtran}

\usepackage[
left=1.9cm,top=1.9cm,right=1.9cm, bottom=2.54cm
]{geometry}

\usepackage{url}
\usepackage{subfigure}
\usepackage{cite}
\usepackage[pdftex]{graphicx}
\usepackage{pifont}
\usepackage{amsfonts}
\usepackage{multirow}
\usepackage{balance}
\usepackage{comment}
\usepackage{epstopdf}
\usepackage{booktabs}
\usepackage{tabularx}
\usepackage{xspace}
\usepackage{mdwtab}
\usepackage{tabularx}
\usepackage{color}
\usepackage{xspace}
\usepackage{graphicx}
\usepackage{grffile}
\usepackage{marvosym}
\usepackage{amsmath}
\usepackage{paralist}
\usepackage{balance}
\usepackage{marvosym}
\usepackage{booktabs}

\graphicspath{{pics/}}

\begin{document}

\title{Statistical Location and Rotation-Aware Beam Search for MillimeterWave Networks}

\author{
Maurizio Rea, Domenico Giustiniano, Guillermo Bielsa, Danilo De Donno, Joerg Widmer 
}

\maketitle

\begin{abstract}
Beam training in dynamic millimeter-wave (mm-wave) networks with mobile devices is highly challenging as 
devices must scan a large angular domain to maintain alignment of their directional antennas under mobility.
Device rotation is particularly challenging, as a handheld device may rotate significantly over a very short period of time, 
causing it to lose the connection to the Access Point (AP) unless the rotation is accompanied by immediate beam realignment.
We study how to maintain the link to a  
mm-wave AP under rotation and without any input from inertial sensors,
exploiting the fact that mm-wave devices will typically be multi-band. 
We present a model that maps Time-of-Flight measurements to rotation and propose a method to infer the rotation speed of the 
mobile 
terminal 
using only measurements from sub-6 GHz WiFi. We also use the same sub-6 GHz WiFi 
system to reduce the angle error
estimate for link establishment, exploiting the spatial geometry of the deployed APs and 
a statistical model that maps the user position's spatial distribution to 
an angle error distribution.
We leverage these findings to
introduce SLASH, a Statistical Location and rotation-Aware beam SearcH algorithm that adaptively narrows the sector search space and 
accelerates both link establishment and maintenance between mm-wave devices. We evaluate SLASH with
experiments conducted indoors with a sub-6 GHz WiFi Time-of-Flight positioning system and a 60-GHz testbed. 
SLASH can increase the data rate by more than 41\% for link establishment and 67\% for link maintenance with
respect to prior work.
\end{abstract}

\IEEEpeerreviewmaketitle

\section{Introduction} \label{sec:intro}

Exact beam alignment of the highly directional antennas of millimeter-wave (mm-wave) communication systems is necessary to 
achieve high data rates or even just a sufficient link margin for 
communication. The need for fast and efficient beam training strategies has stimulated a variety of research studies, both 
theoretical and 
experimental~\cite{mezzav-1, beam-training, 5-abf, 
sur201560, haider2016mobility, patra2016pimrc, 6-hbf, our_icc_paper,Steinmetzer2017,nitsche2015steering}.
There was substantial progress in terms of beam training efficiency compared to the original brute force or (optionally) 
hierarchical 
training of IEEE 802.11ad, for example through compressive beam training approaches \cite{zhu_noncoherent,Steinmetzer2017} that 
only need to scan a subspace of the available antenna beams. Nevertheless, particularly dense deployments with many devices or 
networks with high mobility remain a challenge. In dense networks with small cell sizes, handovers occur frequently, and a device 
may need to beam train with potentially many APs to determine to which APs it has the best link quality. In this case, using 
context information such as mobile position and angular direction from the AP to the device can provide beam steering information 
 to speed up the 
link establishment
without the need
for explicit beam training ~\cite{capone2015context,nitsche2015steering}.
Yet, most WiFi location systems would 
require additional hardware such as gyroscopes to function in the common case of device rotation for maintaining the mm-wave 
link. Systems using angle-of-arrival information can provide relative orientation~\cite{Gjengset:2014,nitsche2015steering},
but require a number of WiFi antennas that are difficult or impossible to integrate on mobile devices.



Motivated by these considerations, we propose to use a sub-6 GHz WiFi 
Time-of-Flight (ToF) positioning system in order to extract statistical context information for reliable mm-wave
link establishment and maintenance. For ToF measurements, only a single WiFi antenna is required and they can thus be computed 
even with mobile and small form-factor devices.
New WiFi transceivers will typically be multi-band, with a sub-6 GHz WiFi chipset complementing the mm-wave interface. 
Given future dense deployments with smaller communication range at mm-wave frequencies, having multiple sub-6 GHz WiFi APs 
in communication range of a device for ranging and positioning will be common. 

We propose to estimate the angle using positioning information derived from several 
independent multiple sub-6 GHz WiFi links, as the exploitation of the spatial geometry of the APs allows 
to reduce the angle estimation error. Most importantly, we build on the insight that rotation of a mobile device affects the 
collected ToF ranges in 
the frequency domain, which can be used to estimate rotation. Our system thus not only provides position but also relative 
rotation 
information, which greatly facilitates beam maintenance and helps to avoid link disruption.


Our contributions are as follows.
\begin{itemize}
\item We present a model to estimate the speed of rotation of a device solely based on 
ToF ranging measurements. 
We then propose a method to infer the amount of rotation to 
apply to the current antenna beam to avoid mm-wave link breaks (Sec.~\ref{sec:rotation}).
\item Given ToF position estimates, we analytically derive a closed-form expression of the statistical angle error 
that describes the angular region that is most likely to contain the dominant angles of arrival (AoAs) and departure (AoD) of the 
mm-wave channel (Sec.~\ref{sec:angle_error_model}).
\item We design SLASH, a statistical beam search strategy based on ToF ranging and positioning 
information 
collected with sub-6 GHz measurements (Sec.~\ref{sec:algorithm}). 
For link establishment, we use the expression of the statistical angle error to 
narrow down the 
sector search space, exploiting the relationship between the quasi-reciprocity of the mm-wave channel and 
the user's position to further speed up the link establishment. 
For link maintenance, we propose a fast strategy to maintain the mm-wave link by tracking both device rotation and 
distances to APs under user 
mobility. 
\item We conduct extensive studies with a 60 GHz testbed and a sub-6 GHz WiFi ranging and positioning system 
(Sec.~\ref{sec:meas_ass}) to validate our approach through measurements in static and mobile scenarios where the device is carried by a 
human.
We compare SLASH both to the 
802.11ad standard and prior work. Our study indicates that SLASH is very effective in increasing the data rate with
respect to prior work
(Sec.~\ref{sec:results}). 
\end{itemize}

\section{Motivation} 
\label{sec:motivation}

%

mm-wave communication supports physical data rates of several Gb/s using highly-directional phased antenna arrays~\cite{1-5G}.
Examples of technology using mm-wave are the IEEE 802.11ad standard for Wireless Local Area Networks (WLANs) in the 
60 GHz band~\cite{11ad} and 5G cellular networks for licensed mm-wave bands~\cite{3GPP}.
The communication between AP and User Equipment  (UE) in mm-wave requires:

{\em Link Establishment}: beam training is needed to find the AoD (at the transmitter) and the AoA (at the receiver) 
to select an antenna sector pair that allows to 
establish communication and maximizes the received power.\footnote{Note that, in this paper, we interchangeably use the terms 
sector and beams.}

{\em Link Maintenance}: after beam training, environment variations and UE mobility and rotation can cause swift changes of the 
link 
quality, and continuous beam adaptation is needed to maintain a high data rate.

However, frequently performing a time-consuming beam training procedure leads to high latency and overhead, which wastes network 
resources and 
deteriorates the system performance.


\subsection{Beam training}\label{sec:11ad_training}
We take as reference the IEEE 802.11ad beam training.\footnote{mm-wave in 5G is under standardization, with beamforming operations
for both data and control planes~\cite{3GPP,3GPP_2}}
The 802.11ad standard handles the beam training procedure via a Sector Level Sweep 
(SLS) strategy to discover the AP-UE sector pair providing the highest received signal strength (RSS). 
SLS comprises two phases: (i) AP SLS and (ii) UE SLS. During the AP SLS phase, the AP exhaustively 
switches 
across all the available sectors -- ideally covering the entire 360$^\circ$ azimuth -- and transmits training frames marked with 
sector identifiers. The UE receives those frames with a quasi-omnidirectional antenna pattern and identifies the AP sector with the highest RSS. The 
same process is repeated in the subsequent UE SLS phase, where the UE trains its sectors and the AP receives quasi-omnidirectionally. 
Once a connection is established, the link quality degradation due to user mobility is handled through either beam 
refinement and tracking, 
or through a full beam training
procedure, that attempt to determine a new combination of beams with improved link quality.


\subsection{mm-wave angle estimation with sub-6 GHz WiFi}
In order to reduce the beam training delay, we investigate how well the AoD/AoA pair providing the highest RSS can be 
estimated by means of sub-6 GHz WiFi technology, when mm-wave and sub-6 GHz WiFi coexist in the same 
multi-band device.\footnote{The sub-6 GHz WiFi radio 
can also serve as a fallback in case no mm-wave link can be established~\cite{zhang2017,3GPP}.}
Previous approaches in this area are affected by two main problems.

\noindent{\bf Inferring angular rotation.} First, UE rotation must be compensated for. Any variation of the angle 
with respect to the spatial reference system must be estimated and the chosen antenna sector adjusted accordingly, otherwise the 
mm-wave link quality would suffer dramatically.
While additional inertial sensors such as compass and gyroscopes can be used for this purpose, they 
require additional hardware and continuous reading of sensor inputs.
Instead, as wireless manufacturers will produce multi-band chipsets, 
we 
investigate in Sec.~\ref{sec:rotation} how 
to extract rotation information from sub-6 GHz
WiFi Time-of-Flight (ToF) ranges.

\noindent{\bf Errors in the angular estimation.} In case of mobile devices, energy and form factor constraints limit the number 
of antennas 
to one or two, even in mobile chipsets of the last generation~\cite{nitsche2015steering,Sur:2016}. Even if APs may be 
equipped with low-order MIMO transceivers in typical form factors, the median error in the angle estimation of 802.11 
frames received by mobile devices with one antenna
is about 20 degrees~\cite{Sen:2013,Gjengset:2014}, i.e., even in this case significant further refinement is necessary to 
ultimately align the antenna beams. This motivates trying alternative approaches that are less demanding in terms of hardware 
requirements. In our system, we use 
ToF position data and APs with one sub-6 GHz antenna to infer the direction of the target device.
However, directly using position information to derive AoA and AoD is likely to 
fail, 
due to the typical location errors inherent in location systems. In fact, as experimentally verified later on 
in Sec.~\ref{subsec:impact_beam_search}, just a few degrees of error in the estimation may result in a strong signal drop of the 
mm-wave link. 
We thus need a methodology that can dynamically 
select the set of beams where the direction of the UE/AP is statistically expected. 

\begin{figure}[t!]
	\centering
	\includegraphics[width=0.7\columnwidth]{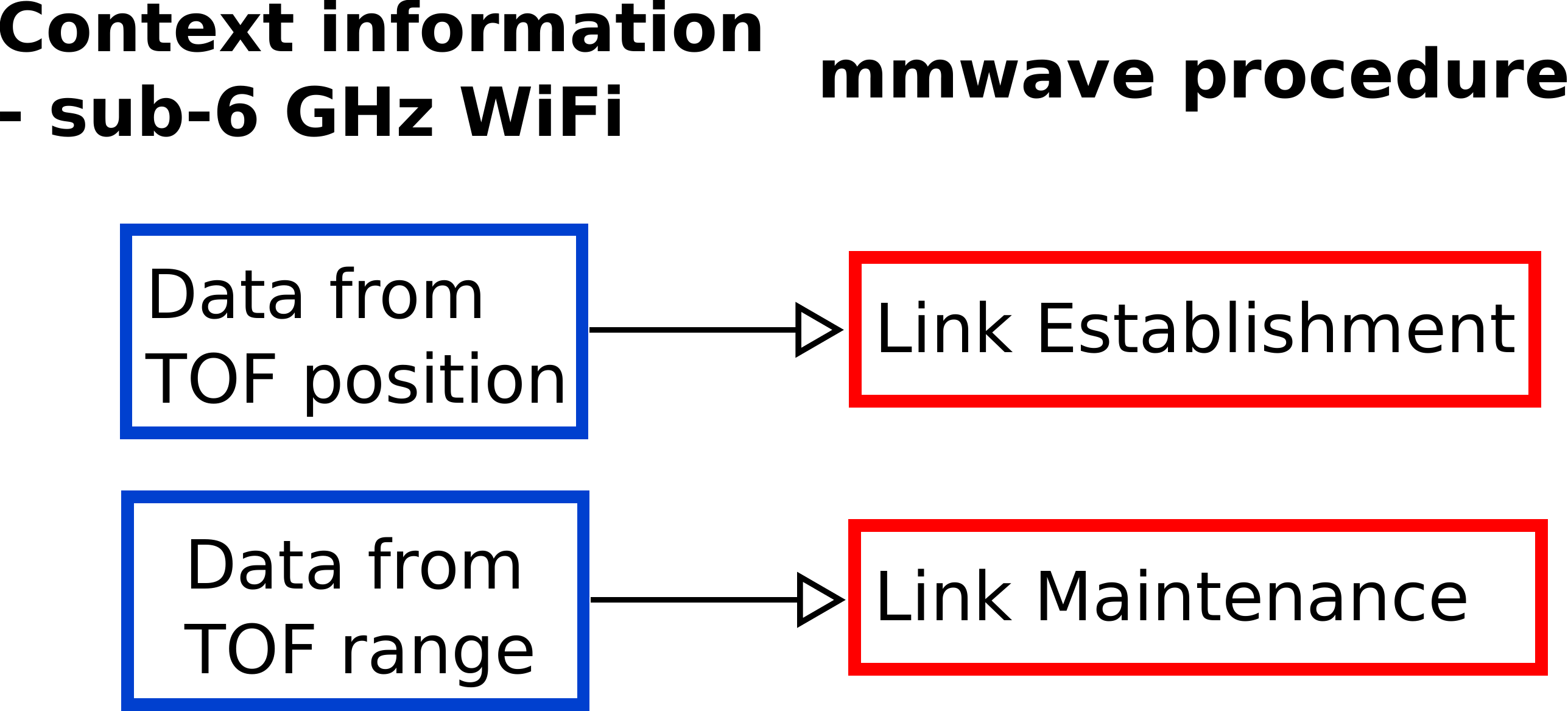}
	 \vspace{-2mm}
	\caption{Context information in SLASH for mm-wave beam search.}
	\vspace{-5mm}
	\label{fig:high_level}
\end{figure}

Based on the above observations, our approach is to design a system that uses commercial 
off-the-shelf sub-6 GHz WiFi devices to infer the angle information based on the 
distance and position estimates provided by ToF measurements. A high-level representation of the context
information used in this work and how it is applied is depicted in Fig.~\ref{fig:high_level}.
\section{From TOF ranging to rotation} \label{sec:rotation}

%
%

The first challenge of our system is that, as soon as the UE rotates relative to the AP, the UE
loses its antenna orientation with respect to the AP and a link maintenance procedure is required.
We investigate 
how to address this problem using {\em only radio measurements} and without using any inertial sensors.
In fact, we aim to avoid any application 
installation in the mobile UE (needed to access inertial sensors) and an increase of battery consumption caused by
the inertial sensors. In addition, we cannot rely on GPS, as it is typically not available indoors, the most challenging 
enviroment for mm-wave beam search.   
Our insight is that a short series of ToF ranging measurements from the sub-6 GHz WiFi interfaces of the APs can provide 
not only the distance to compute the UE position,
but also information about the relative rotation of the UE with respect to the APs. 
We first present the fundamental concepts for ToF ranging used in this work (see
Section~\ref{subsec:tof_pos} for further details). 

\subsection{Time-of-flight ranging}\label{sub:Background}
We use two-way ToF ranging for sub-6 GHz WiFi, which has the advantage of being 802.11 standard-compliant. 
We compute the ToF range using regular frames sent by the APs and acknowledged by the UE via 
802.11 ACKs. Since the time $t$ between each transmitted frame and ACK\footnote{This is specified in the IEEE 802.11 WiFi standard as
the 
Short Inter Frame Symbol (SIFS) duration. The real systematic time can deviate from the nominal value and can be measured 
experimentally.} is deterministic, any additional delay $2 \cdot \delta_\text{ToF}$ can be used to infer the distance between two 
nodes (AP and UE 
nodes in our scenario). It follows that the distance $d_{k}$ of sample $k$ can be computed as 
$d_{k}=c \cdot \delta_{\text{ToF},k}$, where $c$ is the speed of light.
For a given set $\mathcal{K}$ of samples,  the client replying with ACKs adds
uncertainty which can
be of several clock cycles, and that can be approximated as normal distributed~\cite{7932994}.
The effect of multipath is that the resulting distribution is the mixture of normal distributions
(Gaussian Mixture Model (GMM)),
with same standard deviation per distribution. 
Clusters are separated 
using iterative Expectation-Maximization (EM) algorithm on the $\mathcal{K}$ measurements, initialized by
k-means++~\cite{arthur2007k}. Finally, we take the path
with the least positive mean as distance estimate $\hat{d}$.


\subsection{Design of rotation estimator}\label{sub:rotation_model}

\begin{figure}[t!]
	\centering
\includegraphics[width=0.65\linewidth]{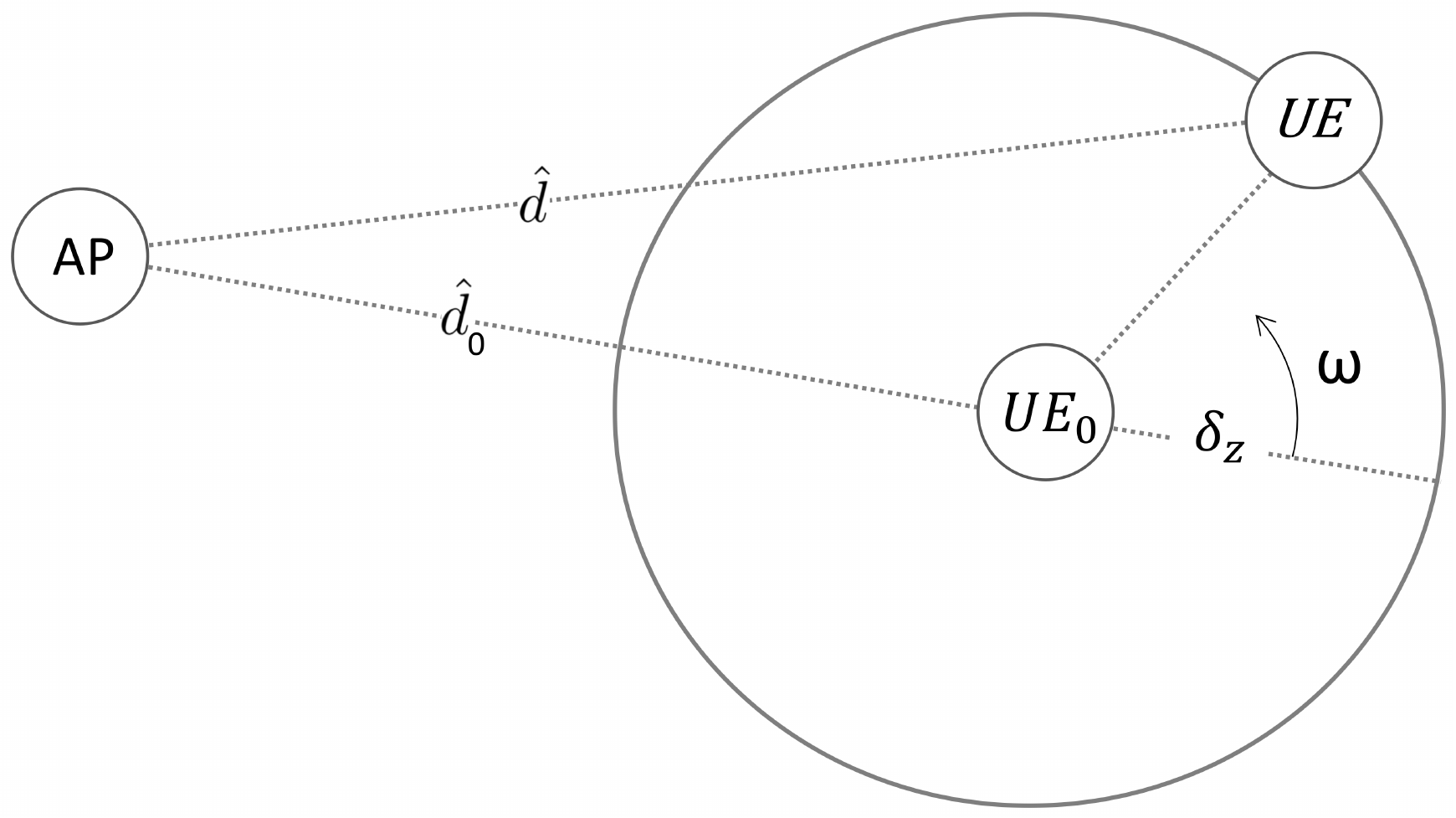}
 \vspace{-2mm}
	\caption{Exploitation of ToF range to define the rotation speed $\omega$.}
	\vspace{-5mm}
	 \label{fig:rotation}
\end{figure}

Let us assume that the rotation has a central point that stays fixed, $UE_0$, and a radius $\delta_z$ which represents the 
distance between the $UE_0$ and its rotation axis.
Referring to Fig.~\ref{fig:rotation}, we can define the relationship between the ToF ranging measurement $\hat{d}$  and 
the rotation speed $\omega$ in polar coordinates applying the theorem of cosine as:

\begin{equation}
{\hat{d}}^2(t)=\hat{d_0}^2+\delta_z^2+2\hat{d_0}\delta_z\cos(\omega t)\, ,
 \label{eq:omega}
\end{equation}
where $\hat{d_0}= \lVert\bold{p}^\text{AP} - \hat{\bold{p}}^\text{$UE_0$}\rVert$.

From Eq.~(\ref{eq:omega}), a $\delta_z$ very close to zero reduces the possibility to estimate the rotation speed. 
However, as the mobile device is carrier by a human, in practice, $\delta_z$ is the distance between the human head and the hand 
(up to the arm length), and thus sufficiently bigger than zero.

\begin{figure}[t!]
	\centering
\includegraphics[width=0.8\linewidth]{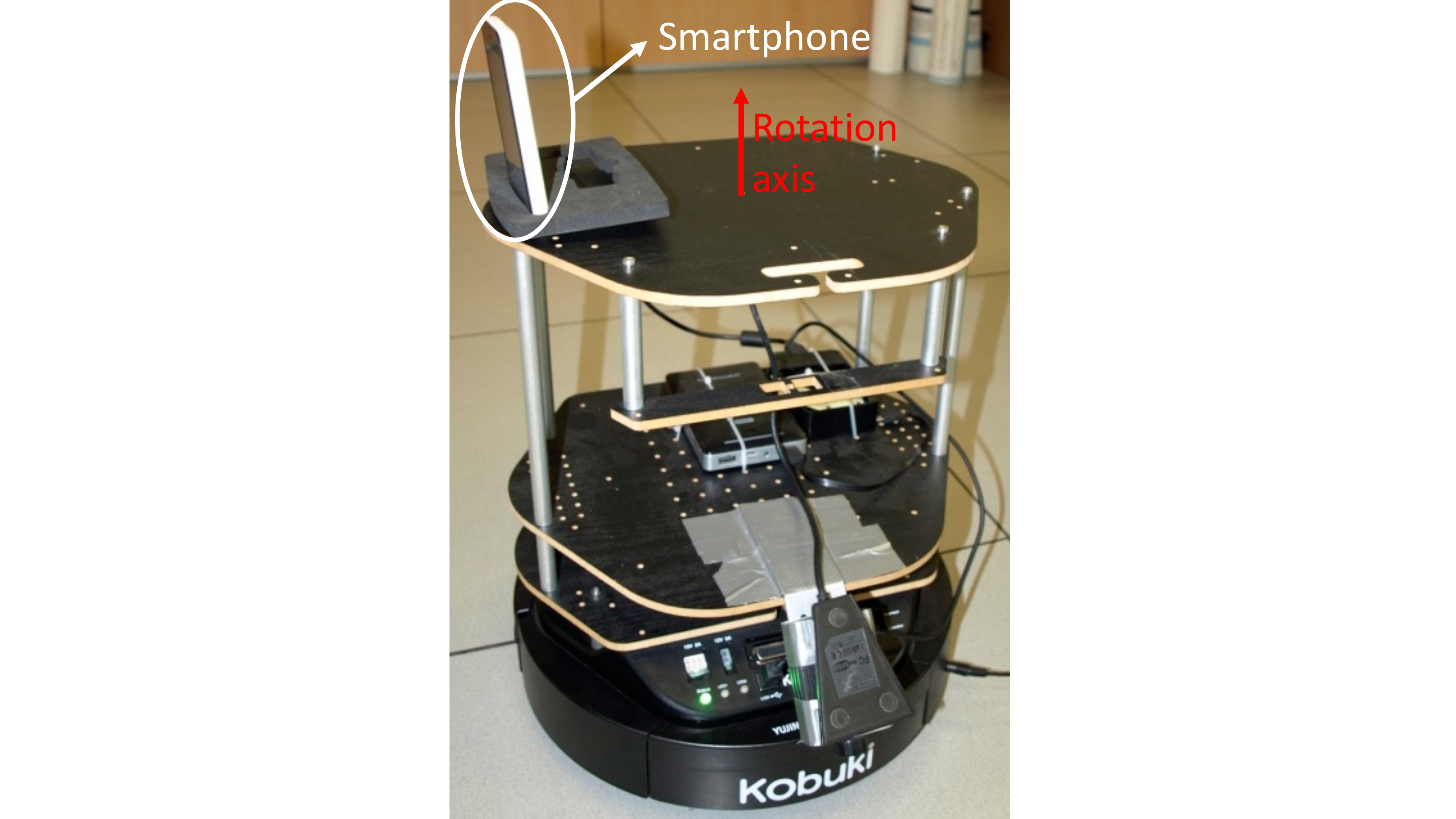}
 \vspace{-2mm}
	\caption{The experimental setup for controlled rotation measurements.}
	\vspace{-5mm}
	 \label{fig:setup_rotation}
\end{figure}

{\bf Dealing with presence of multipath.} 
In order to validate the practicality of this model when using 
real ToF measurements and in presence of multipath, before investigating actual human mobility, 
we first mount a commercial mobile phone as UE on top of a Kobuki Turtlebot II robot running ROS (Robotic 
Operative System), in order to produce a fixed rotation speed. The phone is located at 
5\,m of distance from the AP, in an open area. (The model used for the tests in this 
section is an Alcatel Pixi, but similar results have been achieved with other phones). A picture of the setup is shown in 
Fig.~\ref{fig:setup_rotation}.
Specifically, we consider two tests with the same $UE_0$ 
location, but with $\delta_z$ equal to 0.2\,m and 0.5\,m, respectively. For each experiment, we gather ToF measurements and we 
calculate the estimated distances using 20 samples, according to the methodology in Sec. \ref{sec:meas_ass}. 
For this analysis, we make a new estimation for each collected sample, using a window that selectes the last 20 samples. 
Next, we select only sequences of 20 samples where the number of estimated paths is one.
The reason is that using sequences of samples with more than one path would cause changes to $\hat{d_0}$ 
that are not due only to the rotation.

\begin{figure}[]
	\centering
\includegraphics[width=0.8\linewidth]{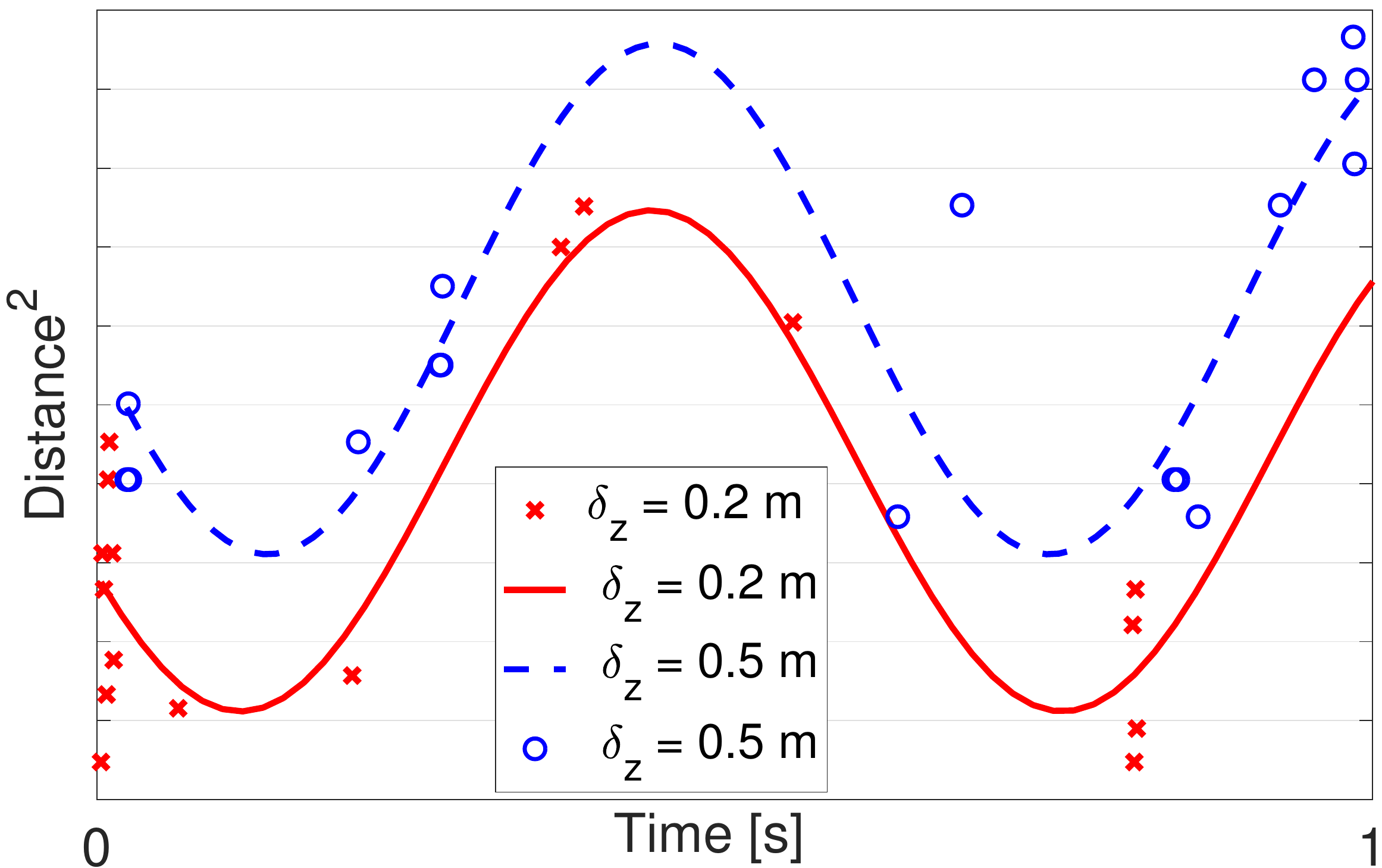}
 \vspace{-2mm}
	\caption{Harmonic fitting of square distances in 1\,s window for $\delta_z$=0.2 m and  $\delta_z$=0.5 m in "ideal" case 
of rotation.}
	\vspace{-5mm}
	 \label{fig:deltaz_difference}
\end{figure}

A key aspect is how to identify the number of dominant paths (clusters) $\kappa$. Samples are distributed according to a GMM 
model, where each cluster identifies one Gaussian component. In a normal indoor environment, the number of 
dominant paths is 
typically up to $5$~\cite{Kotaru15}. However, as we have a 
short train of packets (20 samples) and our ToF system synchronizes only to the strongest path in each sample, fewer clusters are 
expected and we limit the number of clusters to $3$. We infer the optimal $\kappa$ for the GMM statistical model selecting
the model with the lowest Akaike Information 
Criterion (AIC) for the $N$ measurements~\cite{konishi2008information} and select only the sequences with $\kappa=1$ for the 
purpose of estimating the rotation. The Fig.~\ref{fig:deltaz_difference} shows the best fitted harmonic of the square distances 
using least squares, in a 1\,s 
window and using the clusters with $\kappa=1$. 
From the figure, we observe the 
amplitude differences between the two cases, due to the $\delta_z$ variations only.

{\bf Dealing with the remaining noise of the system.} As the remaining multipath noise 
introduces high-frequency components, the result of Eq.~\ref{eq:omega} in a real case of rotation is the sum of different 
harmonics at different 
frequencies. Our analysis shows that the rotation component is located at low frequency and it becomes 
under transformation to the frequency domain. For this study we consider a static case both with and without rotation. In both 
cases, 
AP and UE are positioned at a distance of 5\,m from each other in an open space environment where multipath can occur. For 
rotation, the robot on top of which the UE
is mounted rotates at three speeds: 0.1, 0.35 and 0.7\,rad/sec. 
For each experiment, we gather ranging measurements in a 1\,s window, using the methodology presented above. 

As samples are removed after multipath detection, we 
oversample the sequence to get more accurate estimates. We then interpolate the sequence, by upsampling it by a factor of 
ten and applying a low-pass filter. We compute the power spectral density $P(f)$, normalize it after DC 
removal and we introduce a criterion to estimate the rotation frequency based on the observation that faster rotations have the 
main harmonics at lower frequency. We integrate the normalized power spectral density 
and calculate $f_{1/2}$, the frequency at which the normalized power is equal to 1/2: $\sum_{f_0}^{f_{1/2}} 
\text{P}(f)=\frac{P_{tot}}{2}$,
where $f_0$ is the first frequency component after the DC component and $P_{tot}$ is the total power after DC removal.
We can then estimate the rotation speed as:
\begin{equation}
{\hat{\omega}}=f_{1/2}\cdot 2\pi.
 \label{eq:omega2}
\end{equation}

\begin{figure}[]
	\centering
\includegraphics[width=0.9\linewidth]{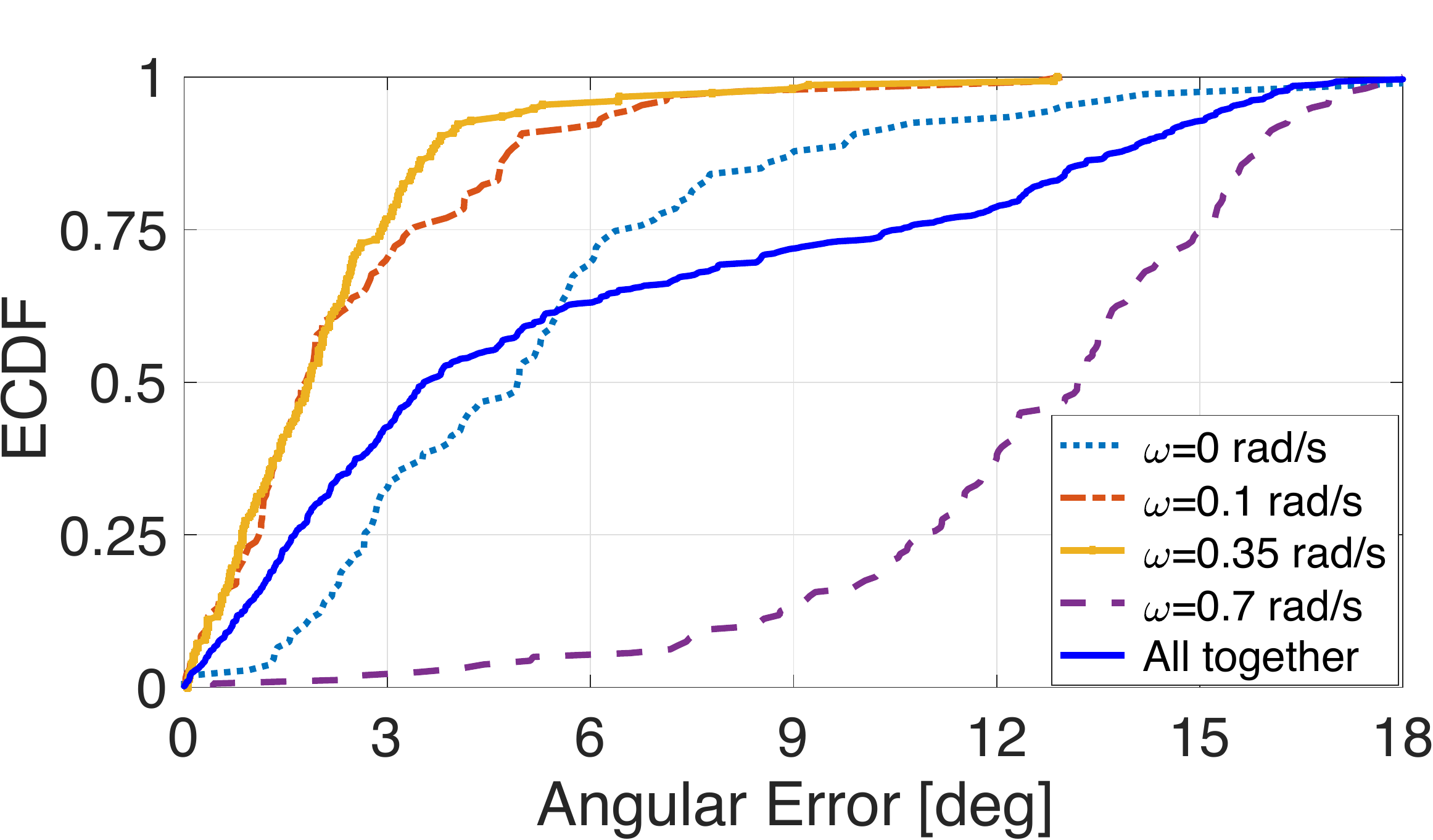}
	\caption{ECDF of the angular error estimating the rotation speed in 1\,s window.}
	\vspace{-2mm}
	 \label{fig:ECDF_omega_estimation}
\end{figure}

It follows that, since we analyze Eq.~(\ref{eq:omega}) in the frequency domain, we are able to estimate the angular 
speed without any knowledge of the radius $\delta_z$, as shown in Eq.~(\ref{eq:omega2}).
We finally evaluate the accuracy of the proposed algorithm for rotation speed estimation.
Fig.~\ref{fig:ECDF_omega_estimation} shows the Empirical Cumulative Distribution Function (ECDF) of the rotation speed error of 
all cases, in degrees.
Due to the noisy measurements, the proposed method has a higher angular error with very small $\omega$, while for high 
$\omega$ (0.7) the method loses in accuracy due to the limited number of collected samples. Yet, as shown in the figure, we are 
able to estimate the rotation rate with a median error 
considering all scenarios of less than 4 degrees in one second (All together in the figure).

\section{Angle error model}\label{sec:angle_error_model}

We derive a statistical model of angle errors from the estimated location, 
spatial geometric information and the desired confidence level. 
 The model will be used in the next sections to estimate where 
the AoD and the AoA in the mm-wave band are statistically expected,
and thus accelerate the mm-wave link establishment.
In our system, we consider indoor deployments with  sub-6 GHz 
WiFi APs 
that can determine the UE position. 

We first present the fundamental concepts for ToF positioning used in this work (see 
Section~\ref{subsec:tof_pos} for further details). 

\begin{figure}
	\begin{center}
\includegraphics[width=0.7\linewidth]{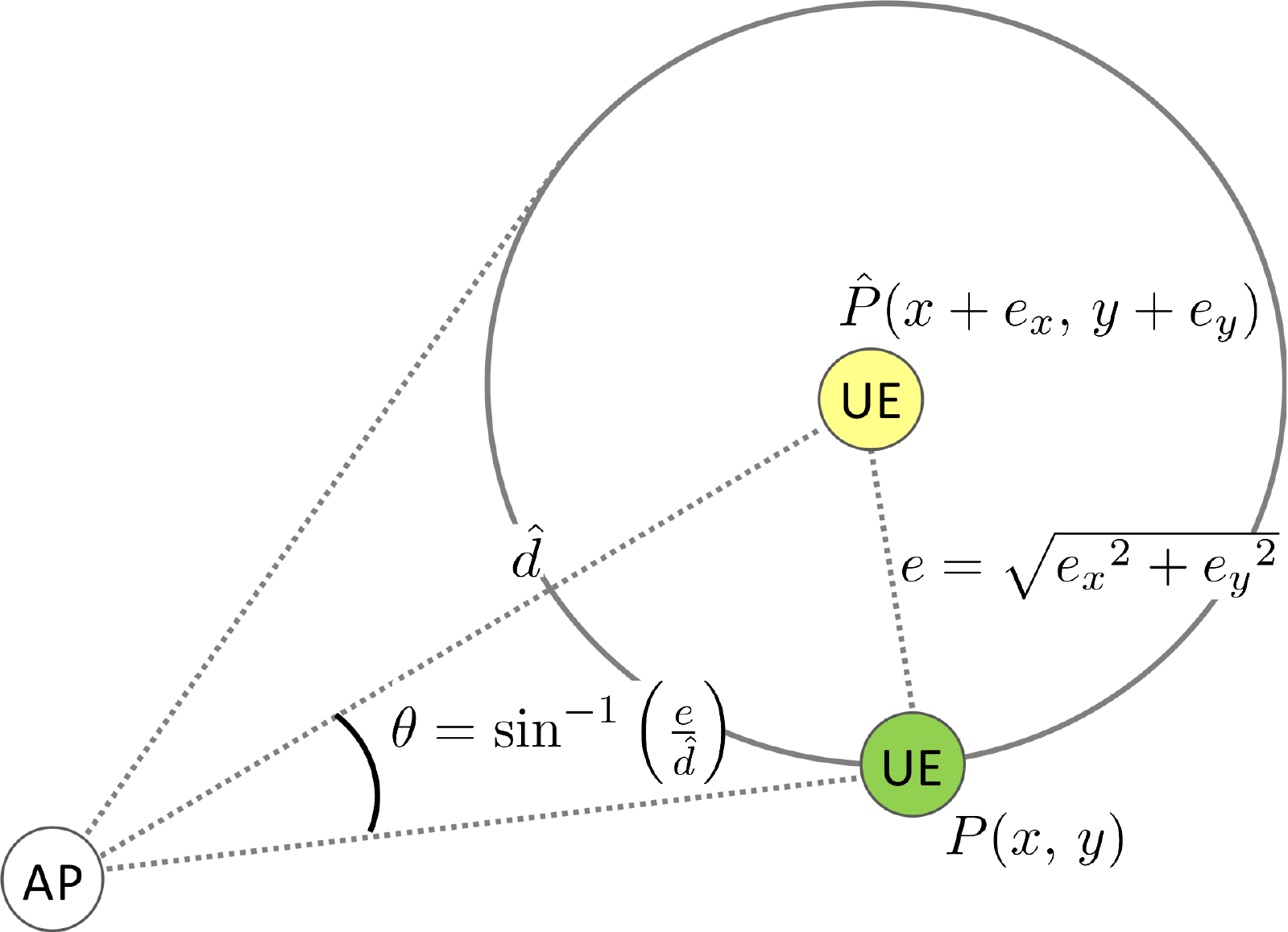}
 \vspace{-2mm}
	\caption{Mapping user position error to angle error.}
	\vspace{-5mm}
	\label{fig:ThetaDefinition}
\end{center}
\end{figure}

\subsection{Time-of-flight positioning}\label{sub:Background_Pos}

Using the distance estimates $\hat{d}$ from at least three different APs to the UE, the UE location 
$\mbox{$\hat{\bold{p}}^\text{UE}$}$ is calculated solving a non-linear least square 
multilateration problem via a weighted Newton-Gauss method. 
Weights are calculated based on the distance error of the new estimated ranges with respect
to the previous computed position.
The positioning system uses a time division scheduler to issue 
measurements rounds where only one AP at a time has the token to measure the ToF 
to the target device. This minimizes hidden nodes, which, based on our experience, can be frequent 
in a method where each AP is free to compete with other APs for accessing the medium. A final interval is 
allocated to drain the measurement queue of all APs. For each measurement round, the order of APs 
changes randomly. 




\subsection{From user position error to angle error}
Let us refer to Fig.~\ref{fig:ThetaDefinition}. We consider a scenario with a fixed AP and a mobile UE. 
We assume a two dimensional Cartesian coordinate system. Extension to the 3D case is straightforward.
The AP position is known and equal to 
$\bold{p}^\text{AP}\in \mathbb{R}^{2\times 1}$. Given \mbox{$\hat{x}=x+e_{x}$} and \mbox{$\hat{y}=$}\mbox{$y+e_{y}$}, the UE's 
real and estimated positions are \mbox{$\bold{p}^\text{UE} = P(x,\, 
y) \in \mathbb{R}^{2\times 1}$} and \mbox{$\hat{\bold{p}}^\text{UE} = \hat{P}(\hat{x},\, \hat{y}) \in \mathbb{R}^{2\times 1}$}, 
respectively. The terms $e_{x}$ and $e_{y}$ represent the 
location 
errors on the $x$- and $y$-axis, respectively. 

\begin{figure}[t]
	\begin{center}
\includegraphics[width=0.7\linewidth]{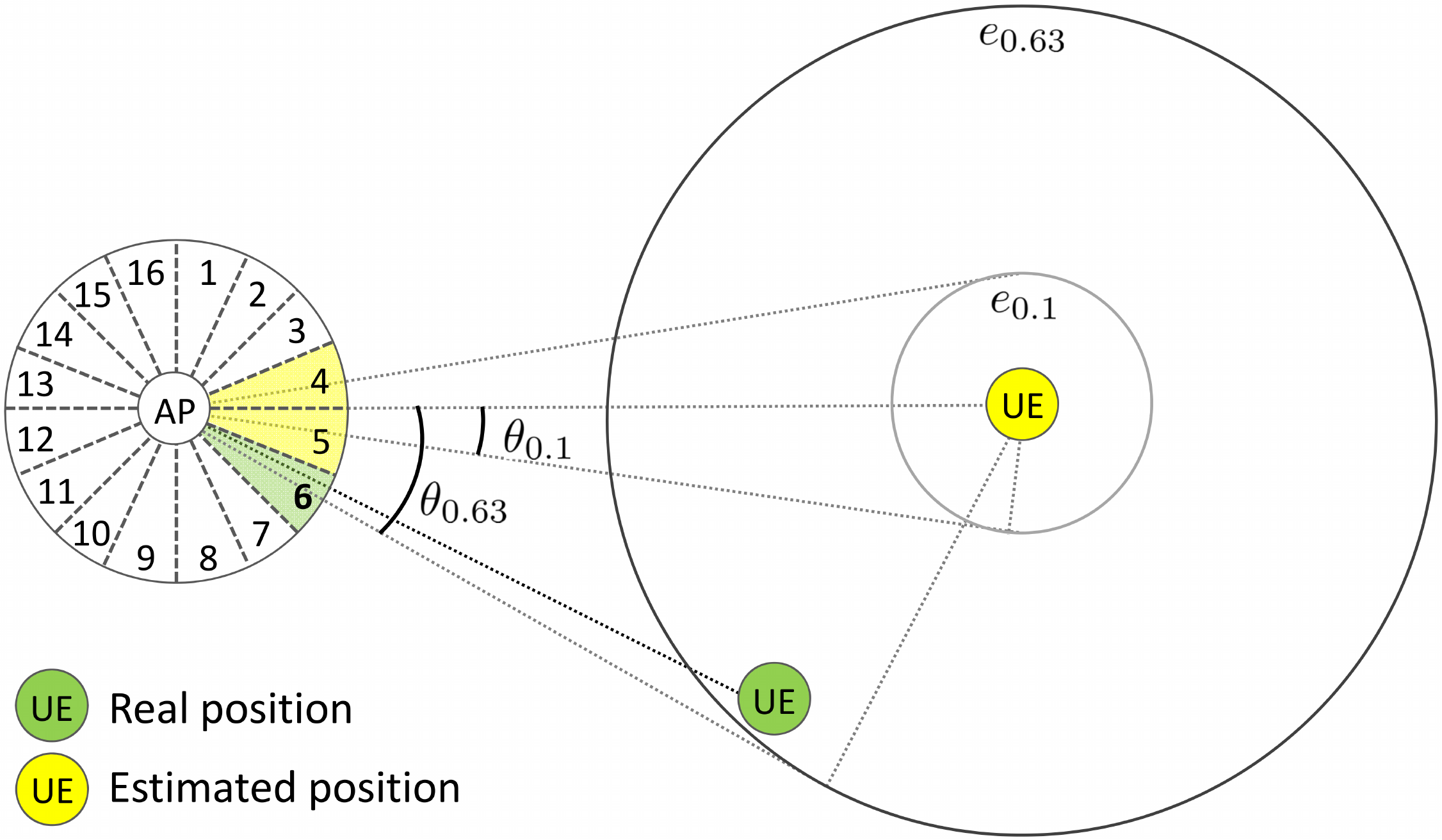}
 \vspace{-2mm}
	\caption{Impact of the choice of two different levels of confidence for the estimated position on the location-aware beam 
	search. As evident, choosing $\theta_{0.1}$ ($p=0.1$) leaves out the optimal beam (green sector) from the subset of 
trained beams 
(yellow 
sectors).}
	\vspace{-5mm}
	\label{fig:ThetaPercentile}
\end{center}
\end{figure}

Let us also define $\hat{d}= \lVert\bold{p}^\text{AP} - \hat{\bold{p}}^\text{UE}\rVert$ as the estimated distance from the AP to 
the UE,
and \mbox{$e=\sqrt{ {e_{x}}^2 + {e_{y}}^2}$} as the UE position error. 
We can then restrict the training sector space to an angular region of width $2\theta$, given by
\begin{equation}
\theta=\sin^{-1}(e/\hat{d})\, .
 \label{eq:theta}
\end{equation}
Referring to Fig.~\ref{fig:ThetaPercentile}, the AP can avoid exhaustive beam search as in the IEEE 802.11ad standard and 
train only a subset of beams (namely Sectors 4, 5, and 6), which contains the sector providing maximum beamforming gain (Sector 
6).

Let us define $p\in[0,1)$ as the level of confidence of the position error. A location error $e_p$ can be defined, which maps, in turn, to an angle error:
\begin{equation}
 \theta_p=\sin^{-1}(e_p/\hat{d})\, ,
 \label{eq:theta_p}
\end{equation} 
which holds for $e_p \le \hat{d}$, i.e., $\theta_p \le \frac{\pi}{2}$ (recall that the training sector space is an angular region 
of width $2\theta$).
From Eq.~\ref{eq:theta_p}, a low $p$ reduces the number of sectors to be trained in the beam search, but it increases the 
probability that the optimal sector is not included in the training. 
An example is shown in Fig.~\ref{fig:ThetaPercentile}. A low $p=0.1$ reduces the number of sectors, but it may not be sufficient 
to find the alignment with the highest RSS. In contrast, a larger confidence level, as such $p=0.63$, results in a larger number of 
sectors, which includes the real UE position and hence the best beam.

\subsection{Derivation of a closed-form expression} 


We derive a closed-form expression of the location angle error as  a function of parameters that can be estimated by the 
positioning infrastructure and of the desired confidence level. Considering ranges with
normal distribution (c.f. Section~\ref{sub:Background}), we can assume that the position 
error is a bi-variate normal 
distribution, where the statistical processes $e_x$ over the $x$-axis and $e_y$ over the $y$-axis have no correlation, and that 
$e_x$ and $e_y$ have zero mean. 

We also define $\mathrm{dRMS}=\sqrt{{\sigma_x}^2+{\sigma_y}^2}$ as the distance root-mean-square 
error, where $\sigma_x^2$ and $\sigma_y^2$, are the variance of $e_x$ and $e_y$, respectively. 
Assuming that the sources of error in the x- and y-axes have the same statistical 
distribution, the statistical processes $e_x$ and $e_y$  have identical normal 
distributions with $\sigma=\sigma_x=\sigma_y$. It follows that $\mathrm{dRMS}=\sqrt{2\sigma^2}$. We can then resort to the 
error modulus 
$e=\sqrt{e_x^2+e_y^2}$ to describe the position error, modeled as a Rayleigh Cumulative Distribution Function (CDF):
\begin{equation} 
F_E(e) = Pr(E \leq e) = 1 - e^{-\frac{e^2}{2\sigma^2}} = 1 - e^{-\frac{e^2}{\mathrm{dRMS}^2}}\,.
\label{eq:error_CDF}  
\end{equation}
The dRMS is often referred to as ``63\% error distance'', meaning that 63\% of errors fall within a circle of radius dRMS, i.e., 
$Pr(E \leq \mathrm{dRMS})\approx0.63$~\cite{kaplan2005understanding}. 

The distance error CDF in Eq. \ref{eq:error_CDF} can be mapped to the location angle error CDF as follows:
\begin{equation} 
\begin{split}
F_\Theta(\theta) &= Pr(\Theta \leq \theta) = Pr(\sin^{-1}(E/\hat{d}) \leq \theta) \\
&=Pr(E \leq \hat{d} \sin{\theta}) = F_E(\hat{d} \sin{\theta})\, ,
\end{split}
\label{eq:F_theta}
\end{equation}
which based on Eq.~\ref{eq:error_CDF} can be expressed as:
\begin{equation} 
F_\Theta(\theta) = 
\begin{cases}
\displaystyle 1 - e^{-\left(\frac{\hat{d}}{\mathrm{dRMS}} \sin{\theta}\right)^2} & \theta \le \frac{\pi}{2} \\ 
1 & \theta > \frac{\pi}{2}\, . 
\end{cases}
\label{eq:theta_CDF}
\end{equation}
We remark that the angular error in Eq.~\ref{eq:theta_CDF} is \emph{not} a Rayleigh function because of the trigonometric function 
$\sin$.

Computing $F_\Theta^{-1}(\theta)$, we can derive a closed-form expression of the location angle error $\theta_p$ as a function of geometric parameters and the desired confidence level $p\in[0,1)$:
\begin{equation}
\theta_p = \sin^{-1}\left(\frac{\mathrm{dRMS}}{\hat{d}} \sqrt{-\ln{(1 - p)}}\right).
\end{equation}

We can remove the dependence on dRMS by resorting 
to the Horizontal Dilution of Precision (HDOP), used in geomatics engineering to measure 
the multiplicative effect of the geometry of the APs on the positioning accuracy based on the (known) AP 
coordinates~\cite{hdop,ogaja2016geomatics}.
The HDOP is computed based on the \emph{unit vector of the 
direction between each AP and the UE}, that is,  $(\bold{p}_n^\text{AP} - \mbox{$\hat{\bold{p}}^\text{UE}$})^T/ \lVert 
\textbf{p}_n^\text{AP}-\hat{\textbf{p}}^\text{UE} \rVert$.
For instance, when the visible APs are in the same line as the UE or close (as it would occur measuring the distance from 
multiple antennas in the same AP), the geometry is unfavorable for positioning and the HDOP value 
is high. In contrast, in the ideal case of perfect spatial geometry (for instance APs distributed in the corners of a square), 
the HDOP 
is close to one for most of UE locations. The formal definition of the HDOP is given in the Appendix.

Using the HDOP, the dRMS can be 
expressed as follows~\cite{kaplan2005understanding}:
\begin{equation}
\mathrm{dRMS} = \mathrm{HDOP} \cdot \sigma_{\hat{d}}\, ,
\label{eq:hdop}
\end{equation}
where $\sigma_{\hat{d}}$ is the standard deviation of the estimated distances for a specific location. Substituting 
Eq.~\ref{eq:hdop} into Eq.~\ref{eq:theta_p} and computing $F_\Theta^{-1}(\theta)$, 
we can derive the following closed-form expression of the angle error $\theta_p$:
\begin{equation}
\theta_p = \sin^{-1}\left(\mathrm{HDOP} \cdot \frac{\sigma_{\hat{d}}}{\hat{d}} \sqrt{-\ln{(1 - p)}}\right)\, .
\label{eq:theta_p_indoor}
\end{equation}

{\bf Estimation process.}  From the above analysis, the estimation process of the angle error $\theta_p$ for a given confidence 
value $p$ with respect to an 
AP, AP$_n$, operates as follows: 
\begin{itemize}
 \item Estimate the UE position \mbox{$\hat{\bold{p}}^\text{UE}$};
 \item Compute the estimated distance $\hat{d}=\lVert \textbf{p}_n^\text{AP}-\hat{\textbf{p}}^\text{UE} \rVert$, where 
 \mbox{$\bold{P}^\text{AP} = \begin{bmatrix} \bold{p}_1^\text{AP} & \bold{p}_2^\text{AP} & \hdots & 
\bold{p}_N^\text{AP}\end{bmatrix} \in \mathbb{R}^{2\times N}$} is the matrix containing the AP coordinates, and 
$\sigma_{\hat{d}}$ the standard 
deviation  over the observation period.
\item Calculate the HDOP based on \mbox{$\bold{P}^\text{AP}$} and \mbox{$\hat{\bold{p}}^\text{UE}$} according to 
Eq.~\ref{eq:hdop_formulaA};
 \item Derive $\theta_p$ based on Eq.~\ref{eq:theta_p_indoor}.
\end{itemize}


\section{SLASH}\label{sec:algorithm}
%

In this section, we present SLASH, a statistical location-aware beam search strategy for mm-wave link establishment and 
maintenance which exploits sub-6 GHz context inputs from the ranges to the APs and from the user position.
For link establishment, we assume that the LOS/quasi-LOS path is the one with the highest RSS between the AP and 
the UE. As shown in recent studies, when LOS/quasi-LOS is blocked, mm-wave 
suffers from outage and significant throughput degradation and the traffic is re-directed to sub-6 GHz 
WiFi~\cite{zhang2017,7959177}. 
For link maintenance, the rotation can be estimated in presence or absence of the mm-wave LOS/quasi-LOS path.





\subsection{Link establishment}\label{sec:slash_le}
\begin{figure}[t]
	\begin{center}
\includegraphics[width=0.85\linewidth]{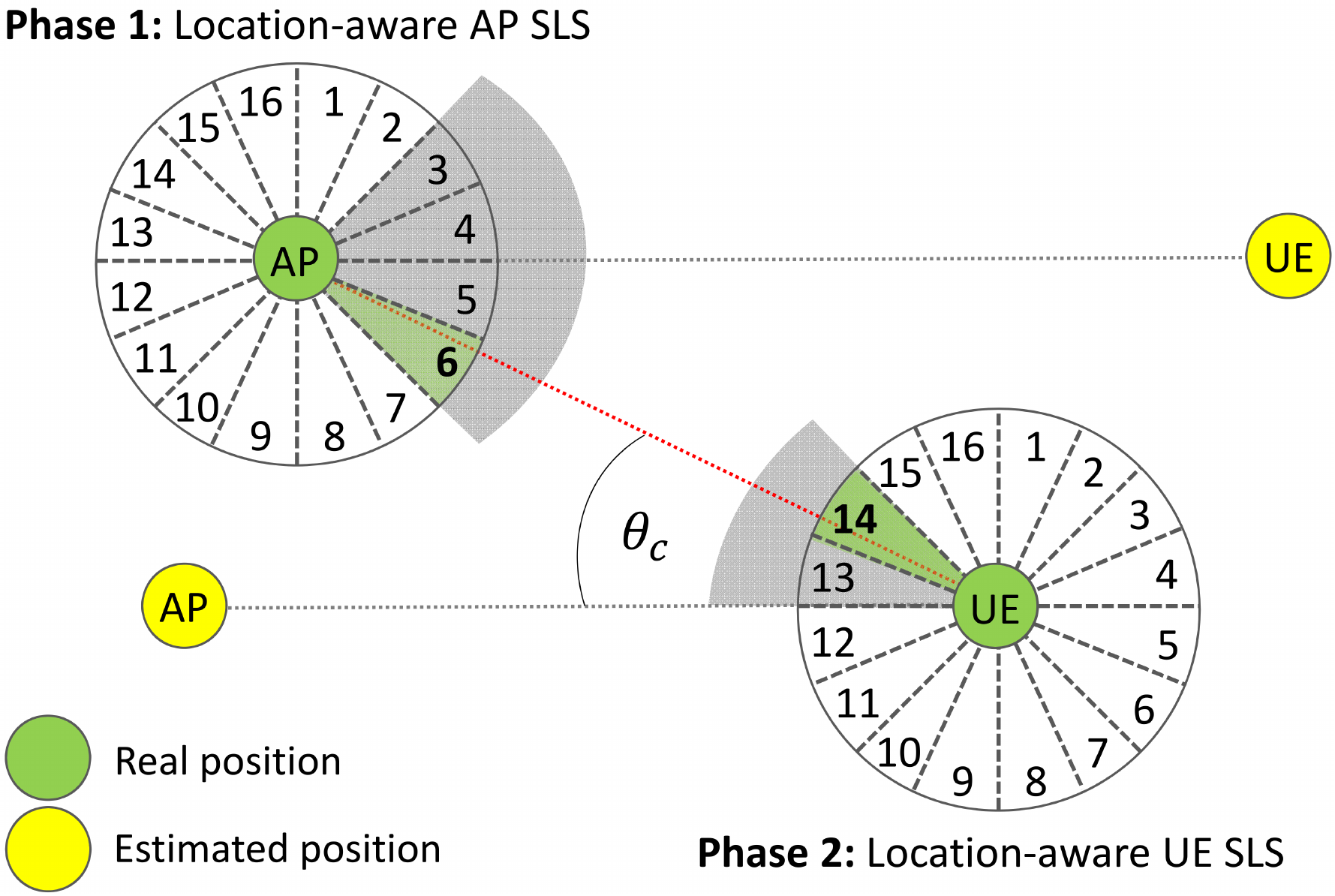}
 \vspace{-2mm}
 \caption{Schematic representation of the working principle for link establishment.}
  \vspace{-5mm}
  \label{fig:ProposedAlgorithm}
\end{center}
\end{figure}

For the link establishment, we propose a beam search algorithm to adaptively narrow the sector search 
space according to the statistical model of 
the angle error developed in Sec.~\ref{sec:angle_error_model}.
We assume that AP and UE employ directional sector antennas of beamwidths $\alpha_\text{AP}$ and $\alpha_\text{UE}$, respectively, 
to communicate at mm-wave frequencies.  We also denote $p_\text{I}$ and $p_\text{II}$ as the desired confidence levels on 
device position used 
during the AP SLS and the UE SLS phases, respectively.
Starting with AP SLS as first stage, the AP can exploit the positioning system to retrieve the UE's estimated position. 
Then, given $p_\text{I}$, the location angle error 
$\theta_{p_\text{I}}$ is computed by the AP according to Eq.~\ref{eq:theta_p_indoor}. The angular portion 
$\Theta_{p_\text{I}}^\text{AP}=2\theta_{p_\text{I}}$, centered around the line joining the AP's real position and the UE's 
estimated position, is used to determine the subset of sectors to probe during \mbox{AP SLS}. 
As shown in the example in~Fig.~\ref{fig:ProposedAlgorithm}, the AP transmits training sequences to probe from beams 3 to 6, 
while the UE receives omnidirectionally and performs RSS measurements.

{\bf Quasi-reciprocity and position of the UE.}
The concept of channel quasi-reciprocity, the similarity of the  
downlink and uplink channels, is usually used for channel estimation~\cite{reciprocity}. We exploit it in a different way:
for the UE SLS phase, we probe the sectors in the direction of the AP (as in the AP SLS phase), but given that the AP SLS already 
resolved the angle error at the AP side, we can take this into account to further reduce the number of sectors in the uplink to 
probe. For the implementation, we set $p_\text{II}=p_\text{I}/2$ for the UE SLS phase. For instance,
considering the example in 
Fig.~\ref{fig:ProposedAlgorithm}, the correcting effect of AP SLS allows to exclude sectors 11 and 12 from the UE SLS, 
leaving just two beams (13 and 14) to be probed instead of four.
At the end of the two phases, the 
AP-UE sector pair providing the highest RSS (beams 6 and 14 in the example) is used for data transmission.

\subsection{Link maintenance}\label{sec:slash_lm}
For the link maintenance, we propose an algorithm to maintain the mm-wave communication and update the current sector using 
ToF measurements both for distance estimation and rotation estimation (Sec.~\ref{sec:rotation}).
As we consider static 
APs, UE rotations imply that only the antenna sector at the UE side needs to be updated. 
The key is that, while the optimal beam for the AP may not change a lot (especially
if the mobile device is not very close to the AP), 
this is not the case with the mobile device, where each rotation can significantly impact the optimal beam.

\begin{figure}[t]
	\centering
	\includegraphics[width=0.85\columnwidth]{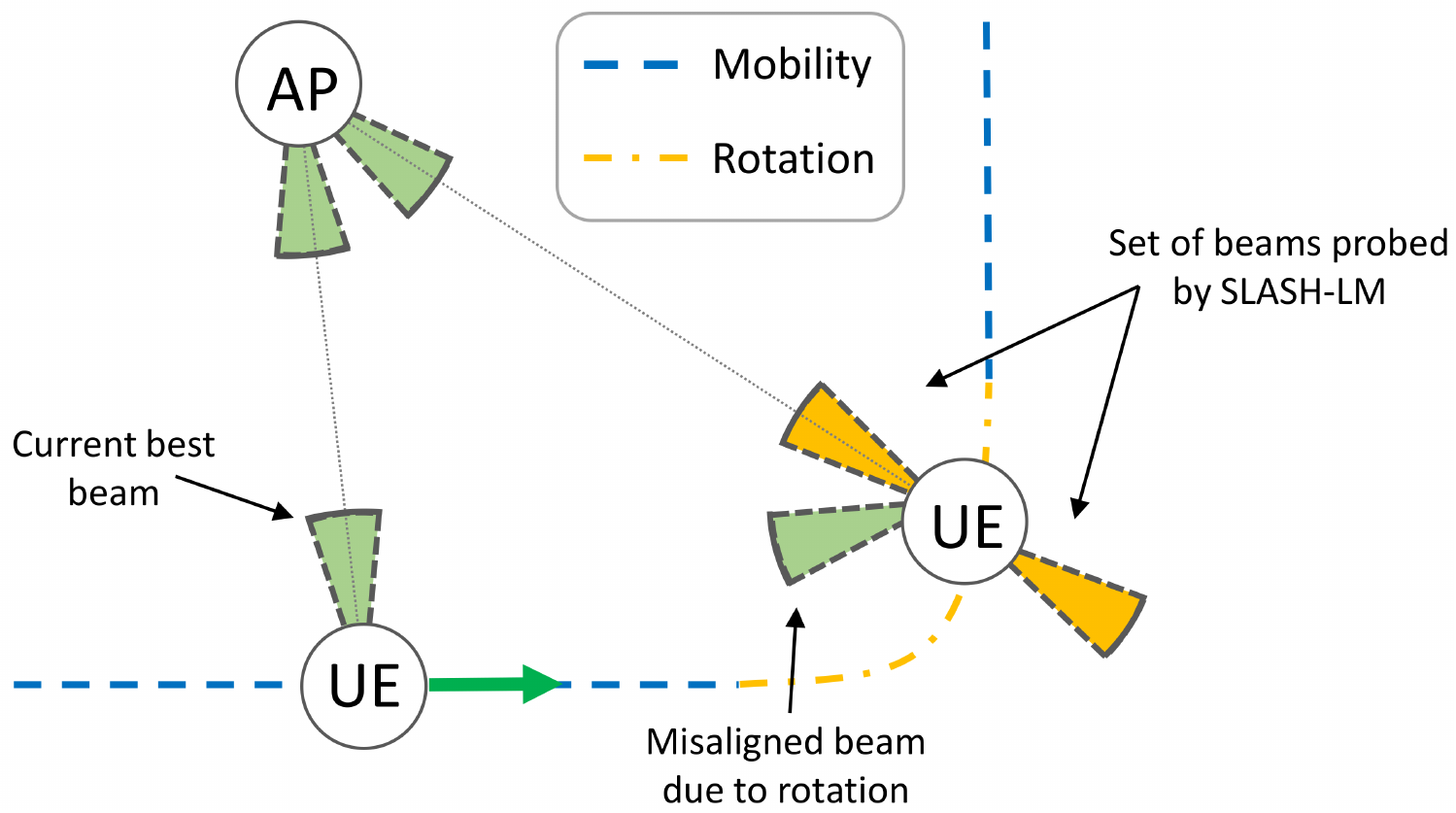}
	 \vspace{-2mm}
	\caption{Illustrative example of link maintenance using SLASH.}
	 \vspace{-5mm}
	\label{fig:slash-lm}
\end{figure}

The main algorithm steps  are as follows:

\begin{itemize}
	\item  During UE mobility, as soon as the system observes a drop in signal quality, the estimated angular rotation 
	since the last link 
establishment or maintenance is determined according 
to the method presented in Sec.~\ref{sec:rotation}; 
	\item If the angular velocity indicates a change in direction, the two candidate sectors are trained following a UE 
SLS-like approach (cf. Fig.~\ref{fig:slash-lm}). Note that there are a total of two 
candidate steering directions at the UEs for typical beamwidths, since our approach does not distinguish between clockwise and 
counterclockwise 
rotations;
\begin{itemize}
 \item  The new direction with the highest RSS is selected for mm-wave communication. Any error in
the estimated angular velocity is immediately solved using the direction with the highest 
mm-wave RSS as feedback loop.
\end{itemize}
	\item If the angular velocity indicates no changes in direction and the mm-wave link quality is low, SLASH monitors the ToF 
distance estimates to the AP. 
\begin{itemize}
 \item A beam refinement that probes the two adjacent sectors is performed when the distance 
estimates indicates that the UE is connected to the closest AP. 
\item A handover procedure is instead triggered (with algorithm for link establishment as in Sec.~\ref{sec:slash_le}) when the 
ToF distance 
estimates indicate that there is a closer AP. 
\end{itemize}

\end{itemize}

\section{Testbed}\label{sec:meas_ass}

\begin{figure}[t!]
	\centering
\includegraphics[width=0.65\linewidth]{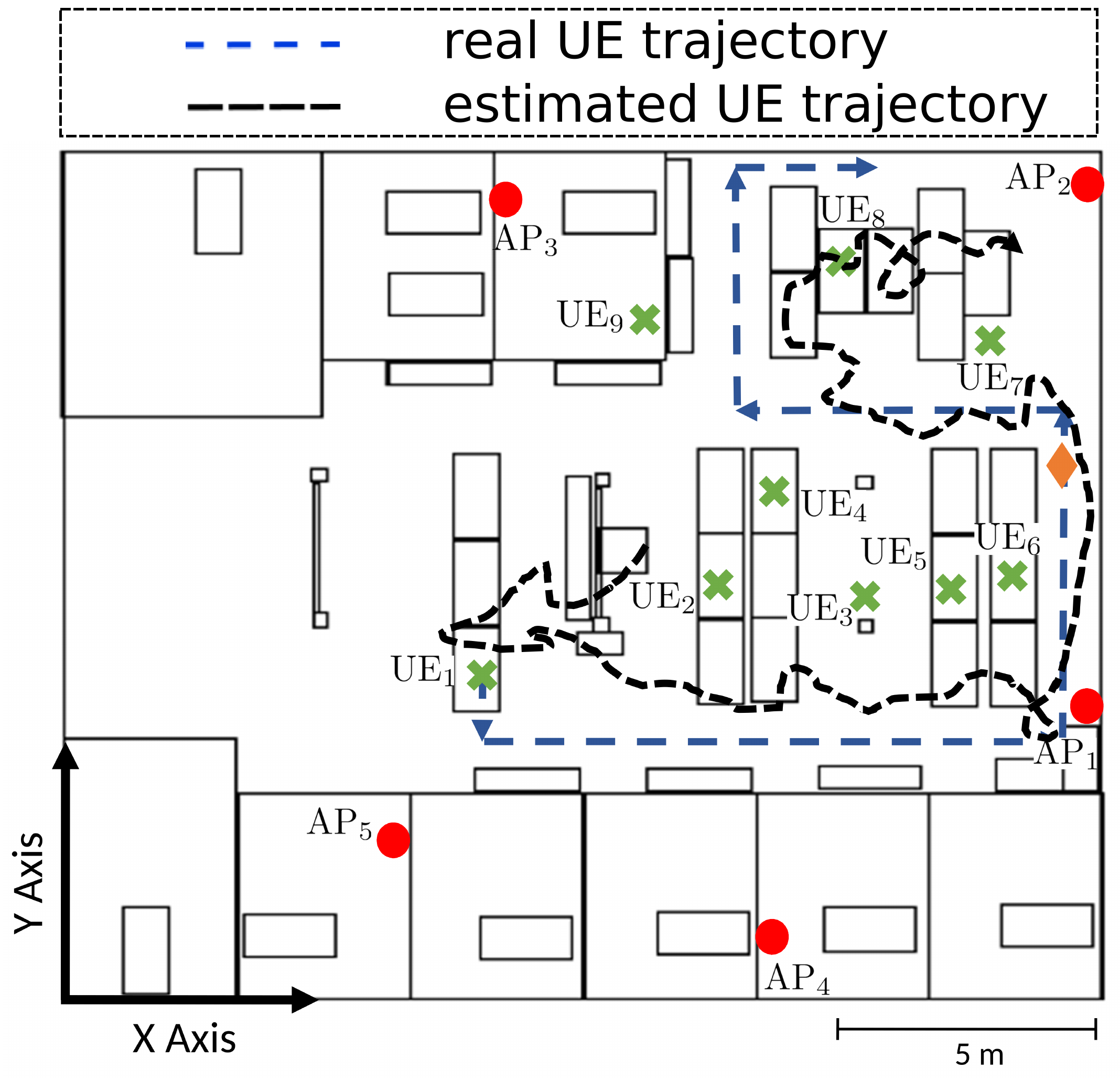}
 \vspace{-2mm}
	\caption{Indoor environment considered for static and mobile tests.}
	\vspace{-2mm}
	 \label{fig:scenario}
\end{figure}

Our measurements are conducted in an indoor space with an open area and offices covering a total
area of 300 m$^2$. Concrete walls separate the offices from the open area, and significant multipath is present in the area.
The map of the scenario is shown in Fig.~\ref{fig:scenario}, where 
red circles mark the five AP positions and green crosses mark nine randomly selected UE positions. 
All APs and UEs are equipped with both 60 GHz and sub-6 GHz WiFi.
In our tests, AP1 and AP2 are used for mm-wave communication as they 
provide coverage in the open area (and the other APs do not), while all APs are used for sub-6 GHz WiFi ranging and positioning.


\subsection{Configuration for Link Establishment}
\begin{figure}[t]
	\centering
	\includegraphics[width=\columnwidth]{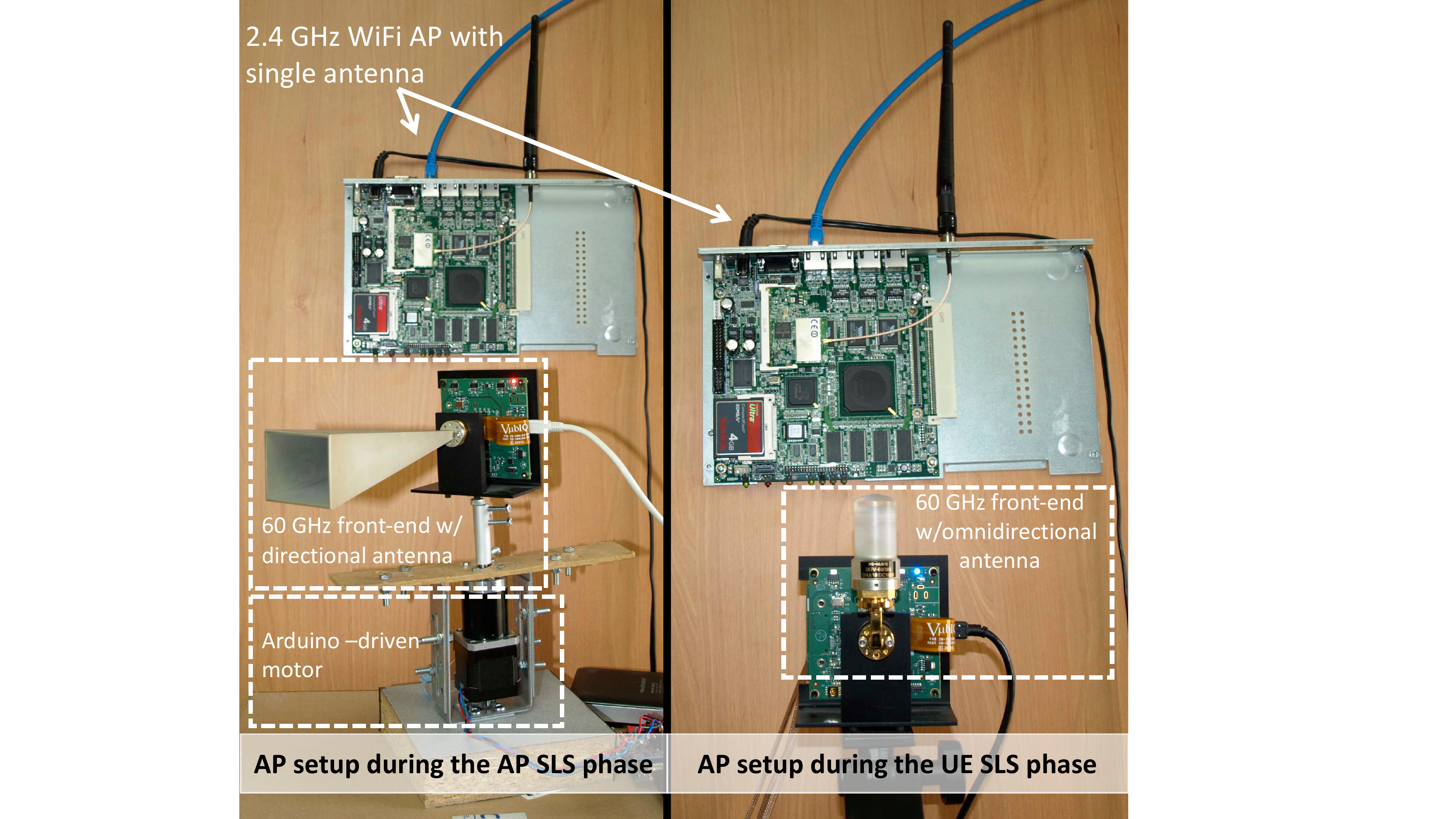}
	 \vspace{-2mm}
	\caption{\mbox{60 GHz} and \mbox{2.4 GHz} configuration of AP during AP SLS and UE SLS phases used for the evaluation.}
	\vspace{-5mm}
	\label{fig:equipment_2}
\end{figure}

Compared to existing works in the literature which limit 
their investigation only to downlink transmissions~\cite{capone2015context, zorzi_eucnc, nitsche2015steering}, 
we consider both the  AP SLS and UE SLS phases.
An illustration of 
the setup of the AP for these experiments is shown in Fig.~\ref{fig:equipment_2}.
In  order to emulate the IEEE 802.11ad AP SLS and the UE SLS link establishment phases, 
in our system, the mm-wave transmitter is connected to an Agilent N5182A signal generator which constantly sends a
probe 
signal, and it is mounted on Arduino-driven stepper motors with an accuracy of 0.18 degrees in order to 
emulate electronically steerable phased antenna arrays. The receiver is connected to an Agilent N9010A signal analyzer to measure 
the RSS. 
In the AP SLS phase, the AP uses a Vubiq 60 GHz transmitter~\cite{vubiq} set to $15$\,dBm and it is equipped 
with a 7$^\circ$-beamwidth horn antenna. The UE uses a Vubiq 60\,GHz receiver connected 
to an 
omni-directional antenna. It has a noise floor equal to $-87$\,dBm. 
The opposite configuration holds in the UE SLS phase. 

\subsection{Configuration for Link Maintenance}

For tests with mobility, 
we have to resort to simulation using a trace-driven approach since our mechanical 
rotation does not allow our system to operate in real-time. Other recent works in mm-wave networks have used a similar 
methodology~\cite{6399486,Sur:2016}.
In this configuration, both the AP and the UE are equipped 
with a 7$^\circ$-beamwidth antenna. Transmission power is set to $15$\,dBm and the noise floor is equal to $-87$\,dBm. 
More specifically, we run 360/7 simulations every step of 25\,cm along the whole trajectory. We then process the 
output data, comparing the real human rotation speed with the estimated one using the 
measurement timestamps.

\begin{figure}[t]
	\centering
	\includegraphics[width=0.75\columnwidth]{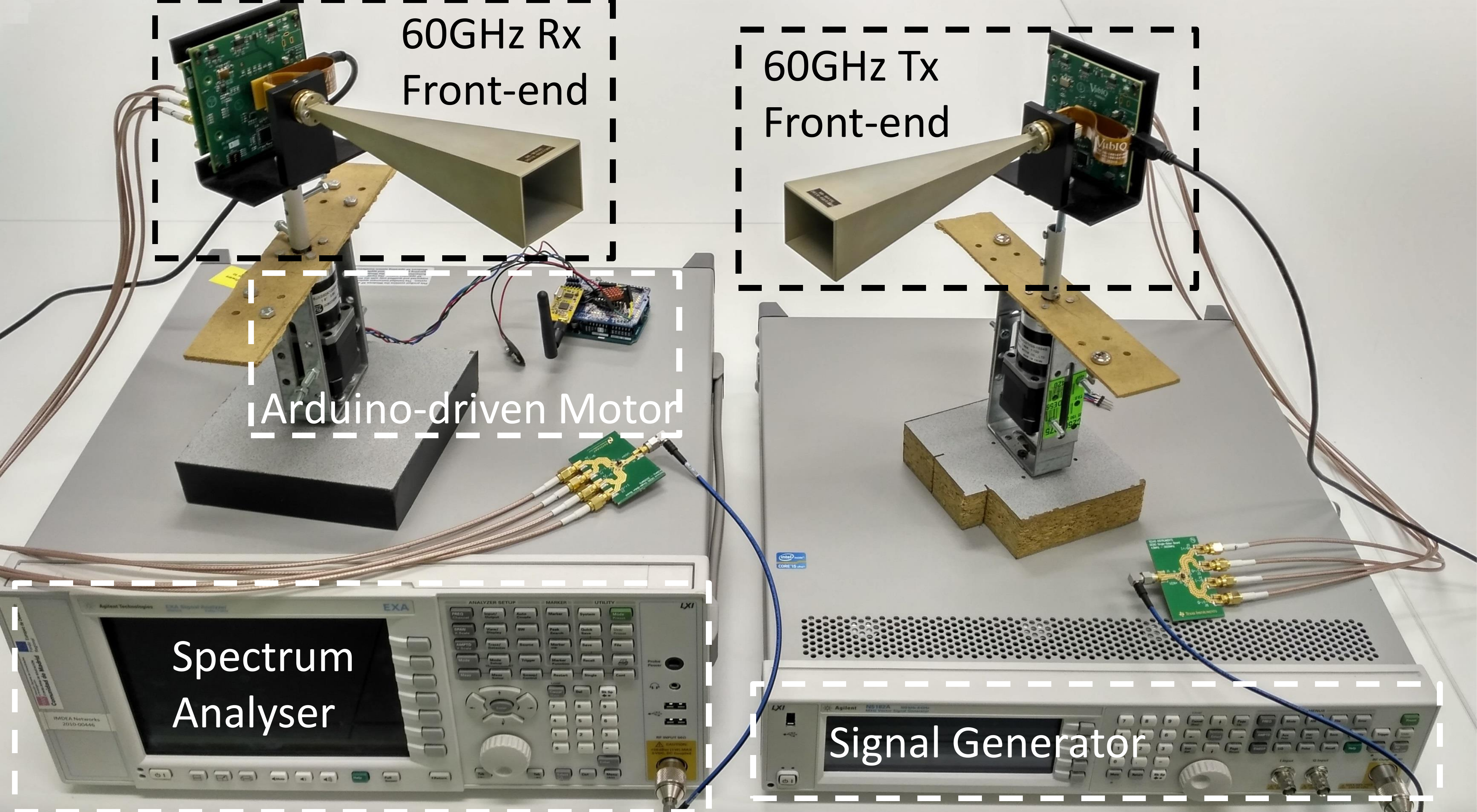}
	 \vspace{-2mm}
	\caption{The experimental setup used for \mbox{60 GHz} RSS measurements once the connection has been established.}
	\vspace{-5mm}
	\label{fig:equipment}
\end{figure}

\subsection{ToF positioning system}\label{subsec:tof_pos}
We design and build a WiFi ToF-based positioning system, where both the AP and the target use a single antenna.
The ToF range is computed using regular 802.11 Probe Responses sent by the APs and acknowledged by the UE via 802.11 ACKs. 
We use Probe Responses rather than normal data because we experimentally observe that the UE replies only to the data packets of its 
associated AP, but not to those other APs. In contrast, the UE replies reliably to the Probe Responses of any AP, which is a 
requirement of our positioning system. In this way, ToF ranges can be computed from multiple APs to estimate the UE 
position. At any point in time, the target device is associated to only one AP, as in typical 802.11 wireless networks.
Our prototype positioning system uses commercial Soekris net5501 embedded machines as APs.
The APs are equipped with an 802.11b/g Broadcom AirForce54G 4318 chipset operating at 2.4\,GHz and one omnidirectional antenna. 
The Broadcom 
chipset runs a 
customized firmware and b43 driver to measure the ToF for each Probe/ACK.  
The APs are connected over Ethernet to the Central Location
Unit (CLU) implemented in C++, that stores and processes the raw ToF data and computes 
the position. The CLU implements also a time-division scheduler to instruct 
the APs to measure the ToF ranges. The order of APs is changed after each scheduling period. The AP 
sends short Probe Response messages of 15 bytes of content and the target replies with 
an 802.11 ACK to these probes. We calculate ranges using 20 samples per AP. 
Tests with the mobile user have been performed by a 
human carrying an unmodified Alcatel Pixi smartphone. 

%
%
%
   
   


\section{System evaluation}\label{sec:results}
%

In this section, we first experimentally validate the model introduced in Section~\ref{sec:angle_error_model}, and then evaluate 
the performance of SLASH in static and mobile scenarios, comparing it against existing solutions in the literature.


\begin{figure}[t]
 \centering
 \subfigure[Model for UE$_1$]
   {\includegraphics[width=0.49\columnwidth]{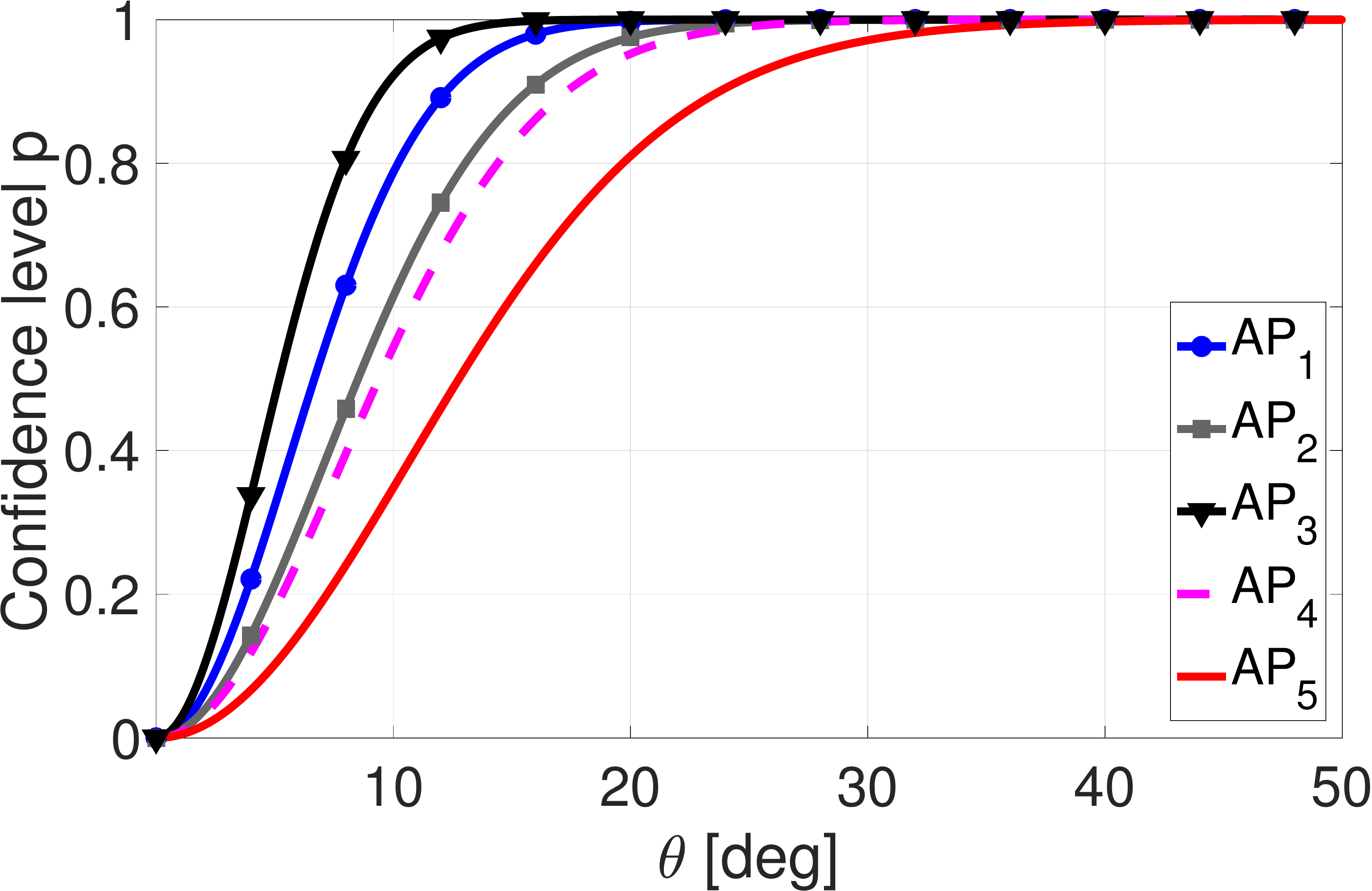}\label{fig:CDF_pos1}}
 \subfigure[Model for UE$_9$]
   {\includegraphics[width=0.49\columnwidth]{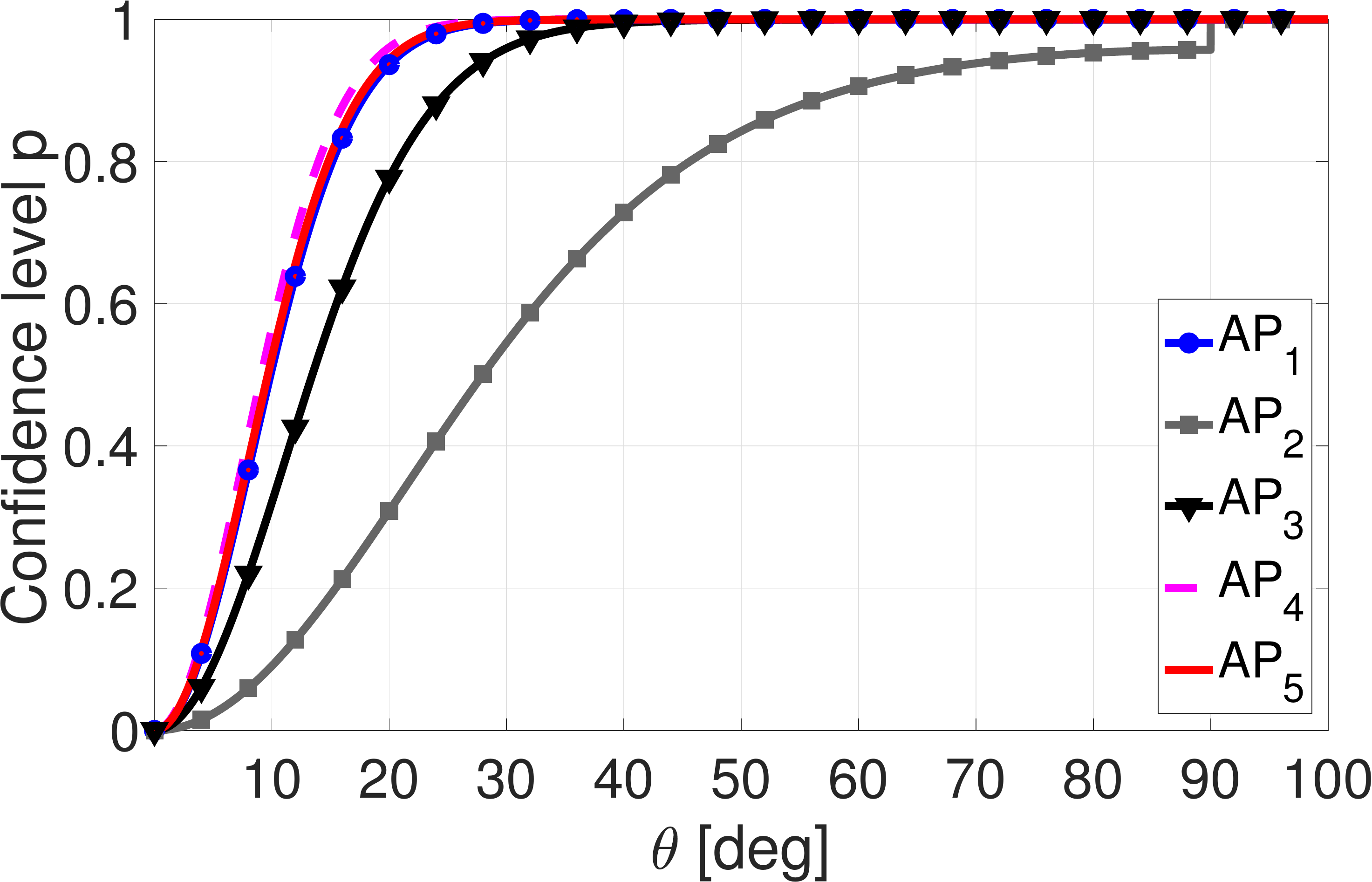}\label{fig:CDF_pos2}}
    \vspace{-2mm}
 \caption{Model CDF of the angle error. }
 \vspace{-5mm}
\label{fig:CDF_experimental}
\end{figure}

\subsection{Model validation}\label{subsec:validation}

We perform extensive WiFi localization measurements in our indoor testbed shown in Fig.~\ref{fig:scenario}.
The median positioning error of our ToF system across all (static) positions is $1.6$\,m. We use this data set to 
analyze the location 
angle error model introduced in Section~\ref{sec:angle_error_model}.
We first show in~Fig.~\ref{fig:CDF_experimental} the CDFs of the angle error at UE$_1$ and 
UE$_9$ locations 
using the model in Eq.~\ref{eq:theta_p_indoor}.
For the study, we use as input i) the estimated AP-UE distance $\hat{d}$ from real experiments, ii) the standard 
deviation $\sigma_{\hat{d}}$ over the observation period, and iii) HDOP computed based on the known position of the five APs. 

The plots in Fig.~\ref{fig:CDF_experimental} confirm the dependence of the CDF on $\hat{d}$ (c.f. Eq.~\ref{eq:theta_CDF}): the 
larger $\hat{d}$, the smaller the angle $\theta$ required to achieve a desired level of confidence. 
The results in Fig.~\ref{fig:CDF_experimental} can be used to obtain $\theta_p$. By selecting on the $y$-axis a 
desired level of confidence $p$ for the position estimate, we can obtain the corresponding $\theta_p$ on the $x$-axis. 


%

\begin{figure}[t!]
	\centering
\includegraphics[width=0.8\linewidth]{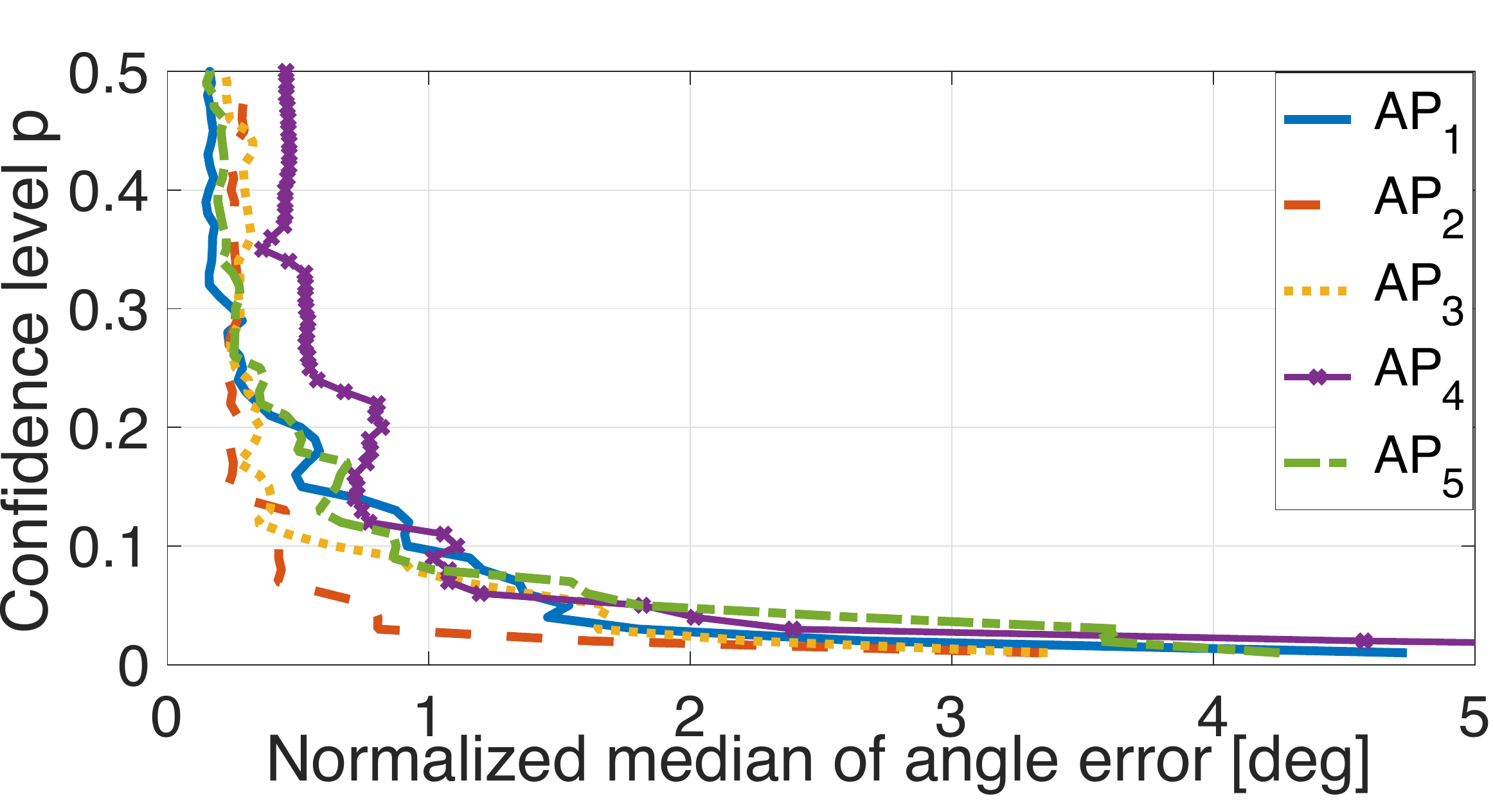}
 \vspace{-2mm}
	\caption{Location angle error for all confidence levels: low error is observed between experimental 
results and theoretical outcomes.}
	\vspace{-5mm}
	 \label{fig:Model_Vs_Real}
\end{figure}

We then assess the validity of this approach for each AP and each $p$, where we compare the 
theoretical values computed using 
Eq.~\ref{eq:theta_p_indoor} (as in Fig.~\ref{fig:CDF_experimental}) with the 
experimental distribution of $\theta_p$. 
Using this data set, we then compute the median of the angle error between the experimental and the theoretical outputs
across all UE positions, normalized with respect to the experimental results,
and plot it in Fig.~\ref{fig:Model_Vs_Real}. 
We can observe that our location angle error model matches very well the experimental findings for a $p$ level higher than 0.1.
In absolute terms, our statistical model provides a median location angle error of $1.44^\circ$ with $p=0.1$ and $4.46^\circ$ 
with $p=0.63$ with respect to the measured one. This shows that multipath in the sub-6 GHz AP-UE links is 
efficiently handled by our ToF ranging and positioning system, and it does not affect significantly the validity of the model. 

%
%

\subsection{Impact of the location error on the RSS for link establishment}\label{subsec:impact_beam_search}

We experimentally investigate how errors in the UE position estimate due to both the limited 
number of samples and multipath that is not resolved by the positioning system affect
the 60 GHz RSS. This study is performed in a few representative positions between the UE and the AP,
using directional mm-wave 
antennas on both nodes (cf. Fig.~\ref{fig:equipment}).
In order to evaluate the link quality for different levels of 
misalignment between AP and UE antenna, we start from an ideal
situation where the nodes are perfectly aligned and measure the RSS when adding different rotation drifts at steps of $7^\circ$ 
(corresponding to the antenna beamwidth). This allows us to reproduce the effect resulting from a misaligned 
mm-wave 
link when the instantaneous output from the localization system is used to steer the devices' antenna beams. 


\begin{figure}[t]
 \centering
 \subfigure[Link 1]
   {\includegraphics[width=0.32\columnwidth]{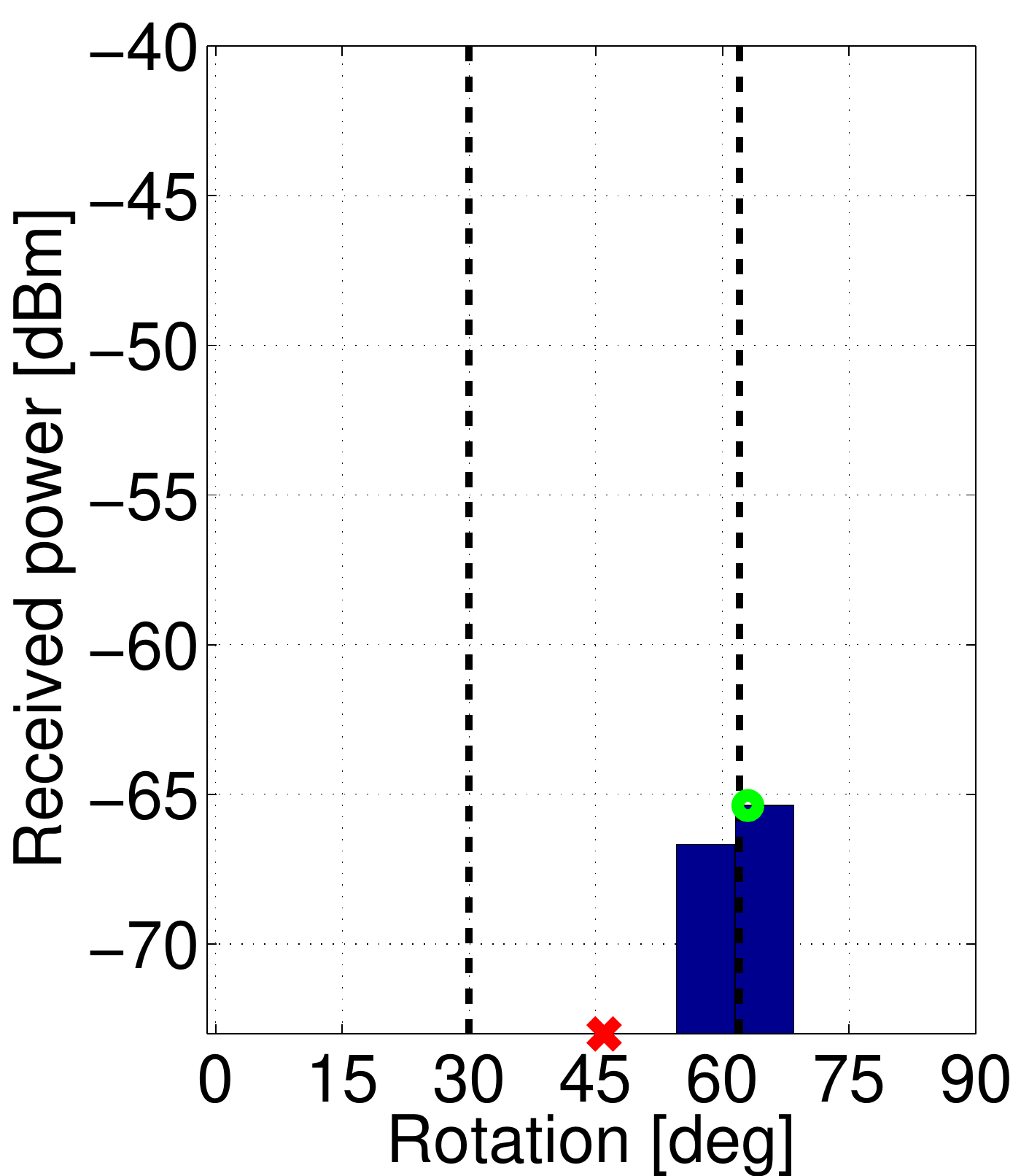}}
 \subfigure[Link 2]
   {\includegraphics[width=0.32\columnwidth]{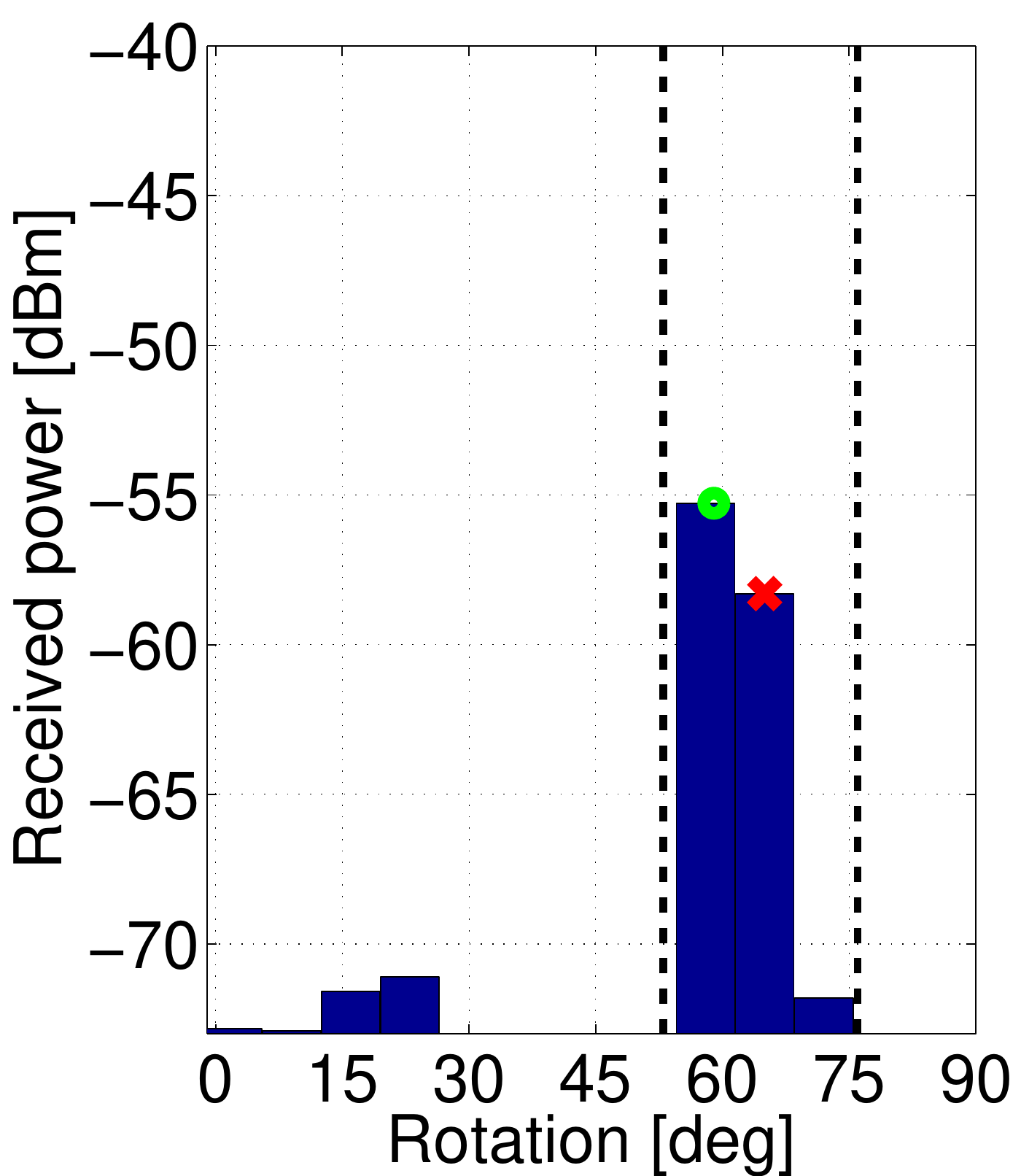}}
	\subfigure[Link 3]
   {\includegraphics[width=0.32\columnwidth]{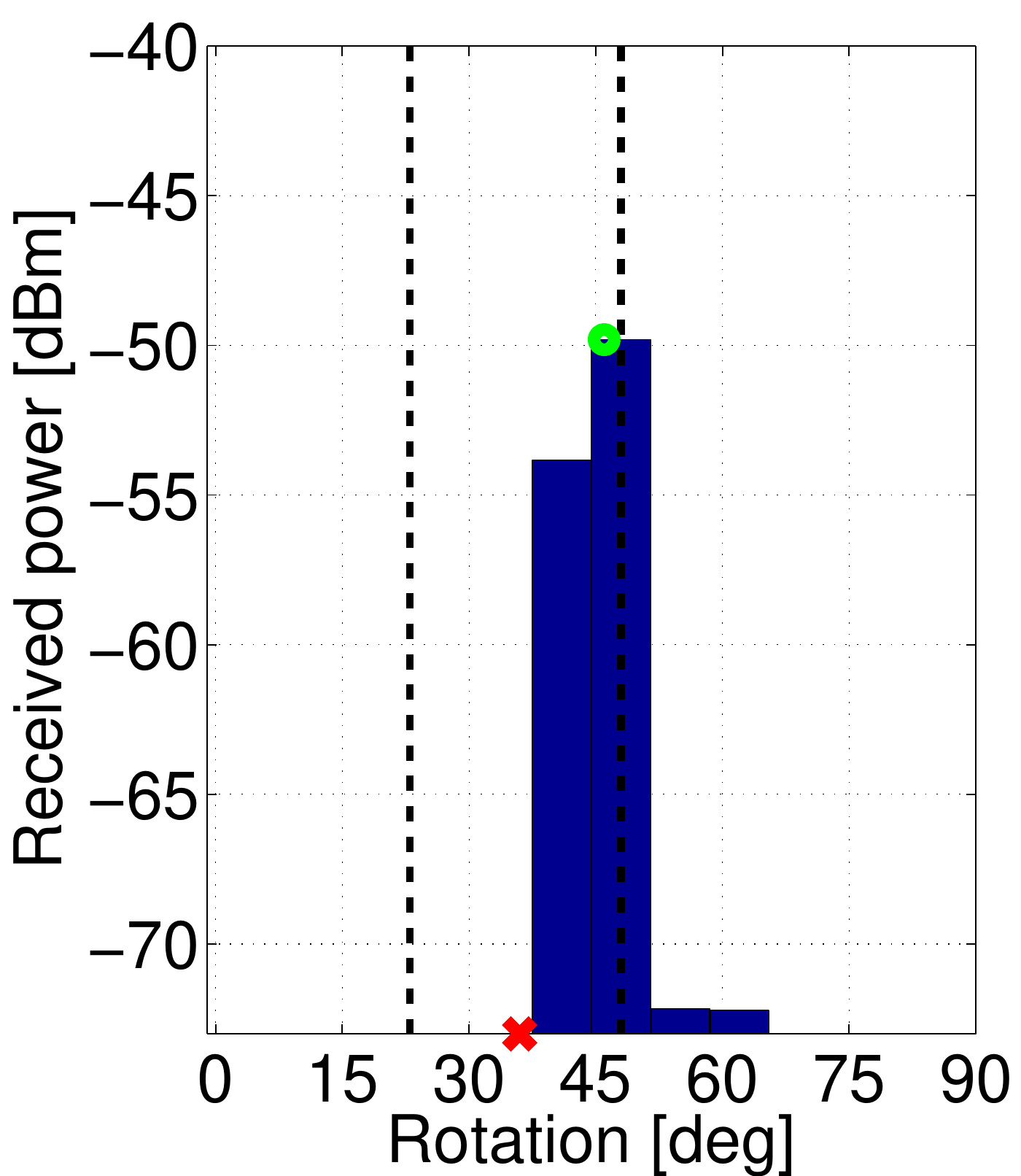}}
	\subfigure[Link 4]
   {\includegraphics[width=0.32\columnwidth]{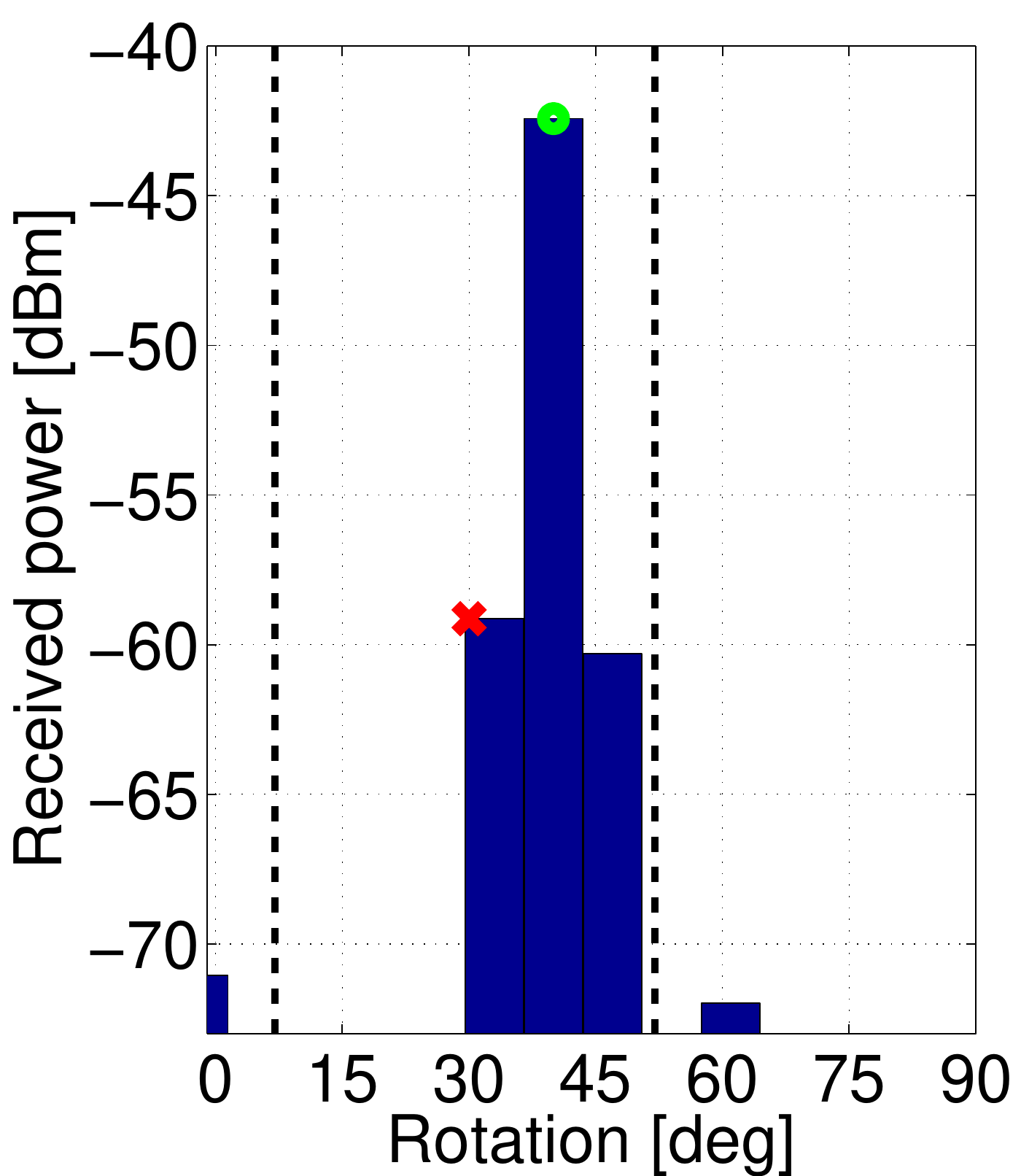}}
	\subfigure[Link 5]
   {\includegraphics[width=0.32\columnwidth]{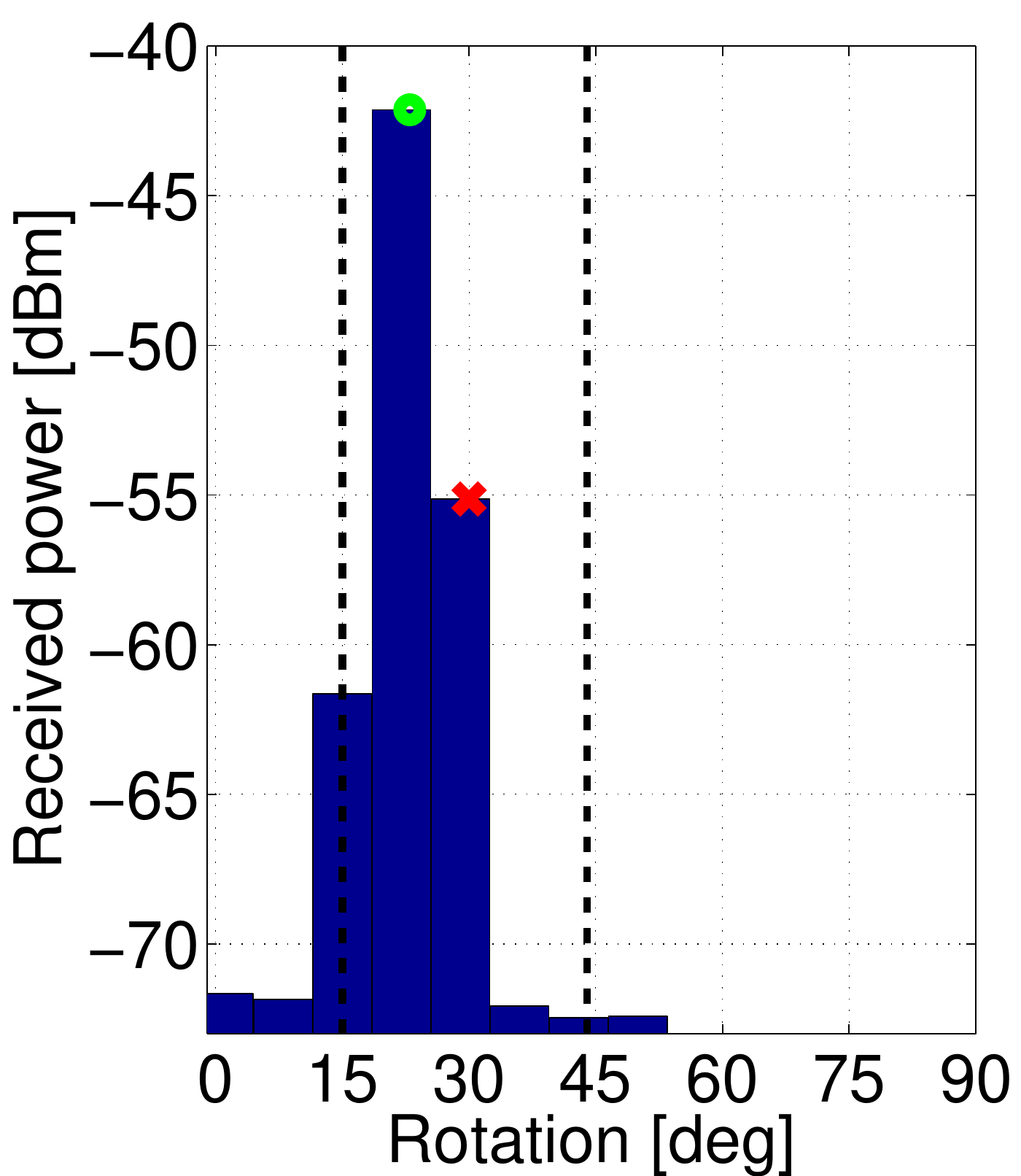}}
	\subfigure[Link 6]
   {\includegraphics[width=0.32\columnwidth]{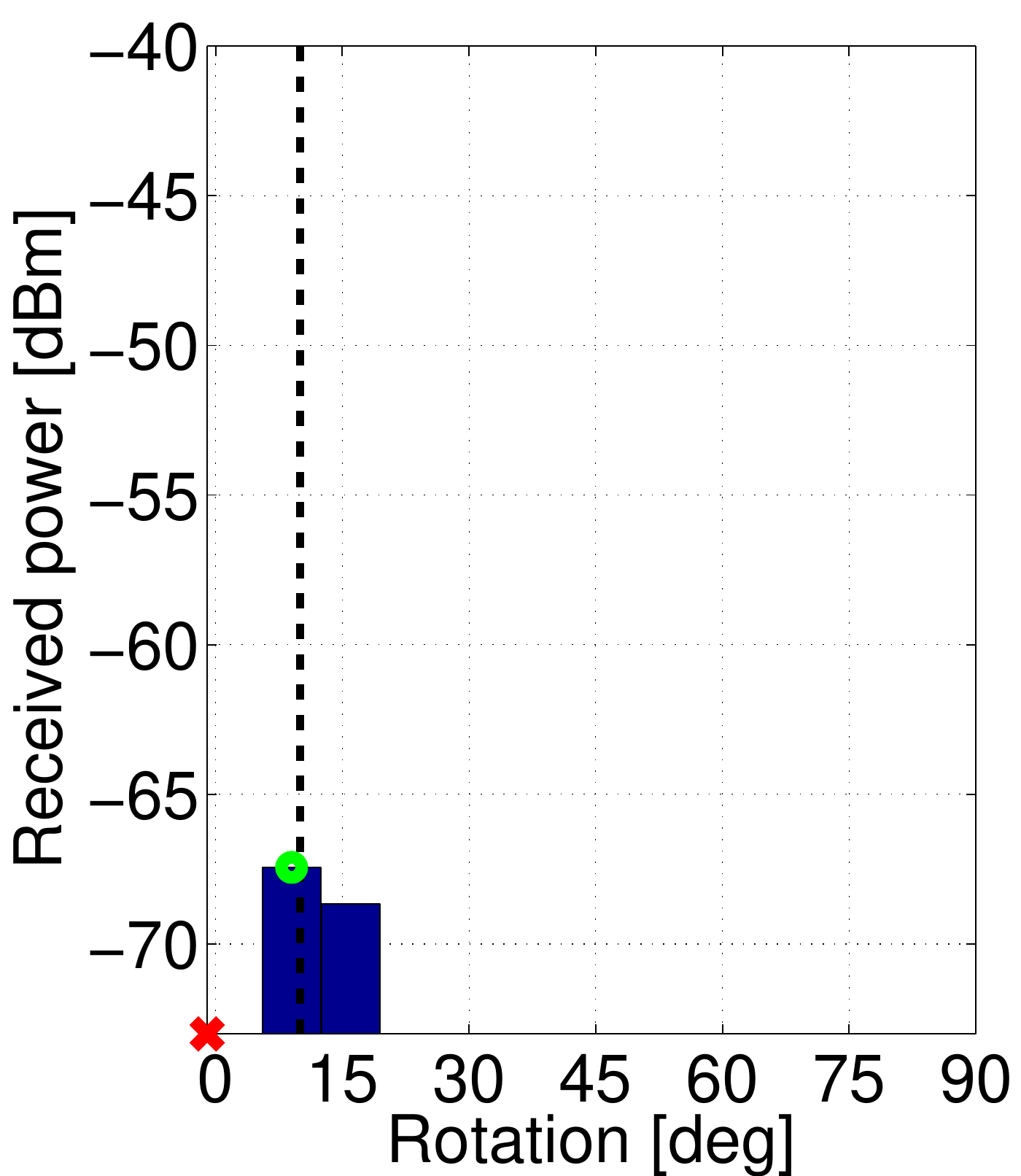}}
    \vspace{-2mm}
 \caption{Measured RSS for different directional AP-UE 60 GHz links when varying the level of misalignment between transmit and 
receive 
antennas.}
\vspace{-5mm}
\label{fig:expresults_snr}
\end{figure}


The results are plotted in Fig.~\ref{fig:expresults_snr}. The 
green circular markers refer to the ideal case with 
perfect knowledge of the AoD/AoA pair (i.e., assuming error-free sub-6 GHz location estimates). In the experiments, we only 
show the RSS values that are above the 
minimum sensitivity level required for correct frame reception. As shown in the figure, even few degrees of 
error in the angle estimation (red crosses) can result in steering in a direction without connectivity.
The dashed black 
vertical lines in the figures represent the angular domain of width $2\theta_{dRMS}$, with respect to the AP position, computed 
applying the location angle error model in Eq.~\ref{eq:theta_p_indoor}. 
From Fig.~\ref{fig:expresults_snr}, we observe that,
for all the links, the highest measured RSS value falls within the angular region defined by $2\theta_{dRMS}$.
The validity of choosing 63\% (dRMS) as a suitable confidence level results will be confirmed in the following section, applying 
the SLASH algorithm for link establishment to narrow down the sector search space.

\subsection{SLASH for a static user}\label{ref:slash_static}

We consider a static user that performs the link establishment phase
through the SLASH algorithm presented in Sec.~\ref{sec:slash_le}.
We use the configuration shown in Fig.~\ref{fig:equipment_2}, and evaluate the ability of SLASH to accelerate the link 
establishment between AP and UE.

\begin{figure}[t]
\begin{center}
\includegraphics[width=0.8\linewidth]{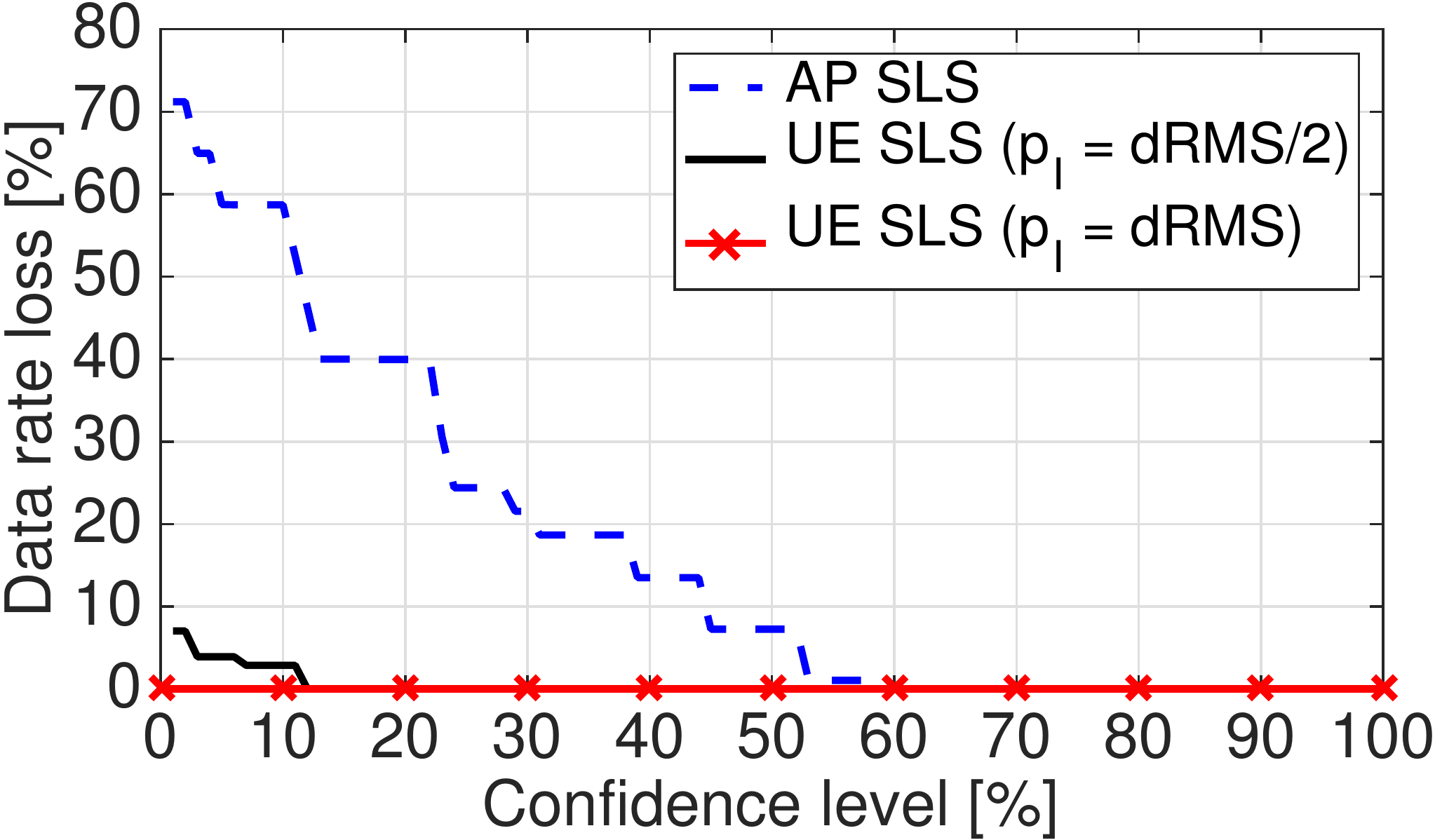}
 \vspace{-2mm}
\caption{Data rate loss for all possible confidence levels during the AP and UE SLS. Loss obtained comparing with respect to the 
standard.}
 \vspace{-5mm}
\label{fig:DataRateLossVSpAP_UE_sls}
\end{center}
\end{figure}

{\bf Data rate loss and confidence level.} As a result of the successful beam search completion, 
the final data rate over the directional 60 GHz link established between AP and UE can be computed for the configured 
beams $B_{best}^\text{AP}$ and $B_{best}^\text{UE}$. More specifically, the experimental RSS values and the noise floor in a 
2-GHz bandwidth at 60 GHz can be translated into an achievable bit-rate $R$ following an IEEE 802.11ad specific rate 
table~\cite{11ad}. 
Using these data, we study the impact of using 
different $p_\text{I}$ (AP SLS phase) and $p_\text{II}$ (UE SLS phase)  values on SLASH performance in a total of 16 AP-UE links. 
Our analysis is 
performed feeding SLASH with the UE position estimates from the sub-6 GHz WiFi ToF positioning system.
We run an 
exhaustive grid search, checking all the possible combinations of such parameters, and compute the data 
rate 
loss with respect to the maximum rate achievable using an exhaustive beam search instead of SLASH.

The results in Fig.~\ref{fig:DataRateLossVSpAP_UE_sls} show that, during the AP SLS phase, setting $p_\text{I}=dRMS=0.63$ is a 
reasonable choice to 
minimize the data rate loss. 
Figure~\ref{fig:DataRateLossVSpAP_UE_sls} highlights also the benefit of exploiting the quasi-reciprocity of the mm-wave channel 
in SLASH as presented in~Sec.~\ref{sec:algorithm}. As shown in the plot, fixing $p_\text{I}$=dRMS, 
the data rate loss for the UE SLS phase reaches its minimum even with the minimum confidence level. In this case we could be able 
to estimate the best beam just updating the steering direction. 
In the same figure, we also show the case $p_\text{I}$=dRMS/2, which results in a small data rate loss for small confidence 
levels of 
the UE SLS phase.

\begin{figure}[t]
\begin{center}
\includegraphics[width=1\linewidth]{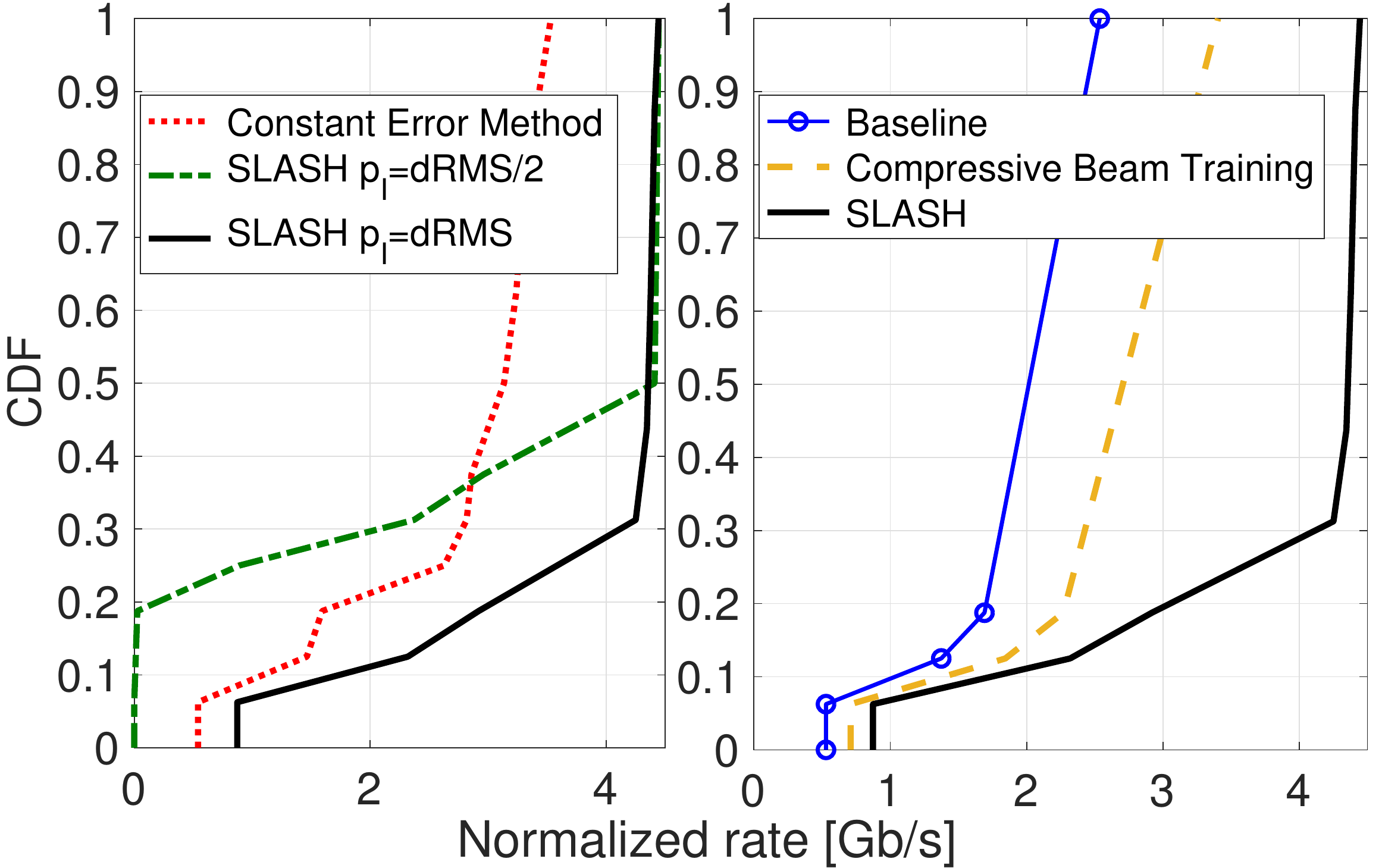}
 \vspace{-2mm}
\caption{ECDF of the normalized mm-wave data rate for different 
beam search strategies.}
 \vspace{-5mm}
\label{fig:ECDF_NormalizedPhyRate_DiffAlgorithms}
\end{center}
\end{figure}

{\bf Normalized rate of SLASH for link establishment.} We then compute the normalized data rate as $R \cdot \frac{T}{T+\tau}$, 
where 
$R$ is 
the achieved data rate, $T$=2 ms is the data frame size, and $\tau$ is the beam search latency. In order to measure the beam 
search latency, we consider the typical duration of training packets according to the 802.11ad standard (15.8 $\mu s$). 
Fig.~\ref{fig:ECDF_NormalizedPhyRate_DiffAlgorithms} shows the CDF
of the normalized data rate for different beam search strategies. 
To further highlight the benefit of exploiting the quasi-reciprocity of the mm-wave channel, 
we 
show, on the left, the performance exhibited by SLASH both with $p_I = dRMS$ and $p_I = dRMS/2$, and compare it against the 
strategy in~\cite{zorzi_eucnc} (``Constant Error Method'')
that uses a constant error for the UE position without developing any angle error model. 
Choosing $p_I = 
dRMS/2$ we can 
increase the average of the speed-up, at the cost of higher probability of falling back to sub-6 GHz frequencies for 
communication. 
SLASH with $p_I = dRMS$ shows the best performance, and 41\% higher in median  than ``Constant Error Method''. We compare 
SLASH with $p_I = dRMS$ with other two different strategies on the right of Fig.~\ref{fig:ECDF_NormalizedPhyRate_DiffAlgorithms}, 
the IEEE 802.11ad as ``Baseline'', 
and~\cite{Steinmetzer2017} (``Compressive Beam Training'').
''Compressive Beam Training`` adapts 
compressive path tracking for sector selection. 
SLASH achieves a data rate that is 64\% higher in median than 
''Compressive Beam Training``. In practical cases, SLASH can achieve even higher performance than prior work,
as it can exploit the knowledge of the estimated distance to APs in range to select the best moment 
in time for handover and performing link re-establishment phase (cf. Sec.~\ref{sec:slash_lm}).

\begin{figure}[th!]
 \centering
   \includegraphics[width=0.91\linewidth]{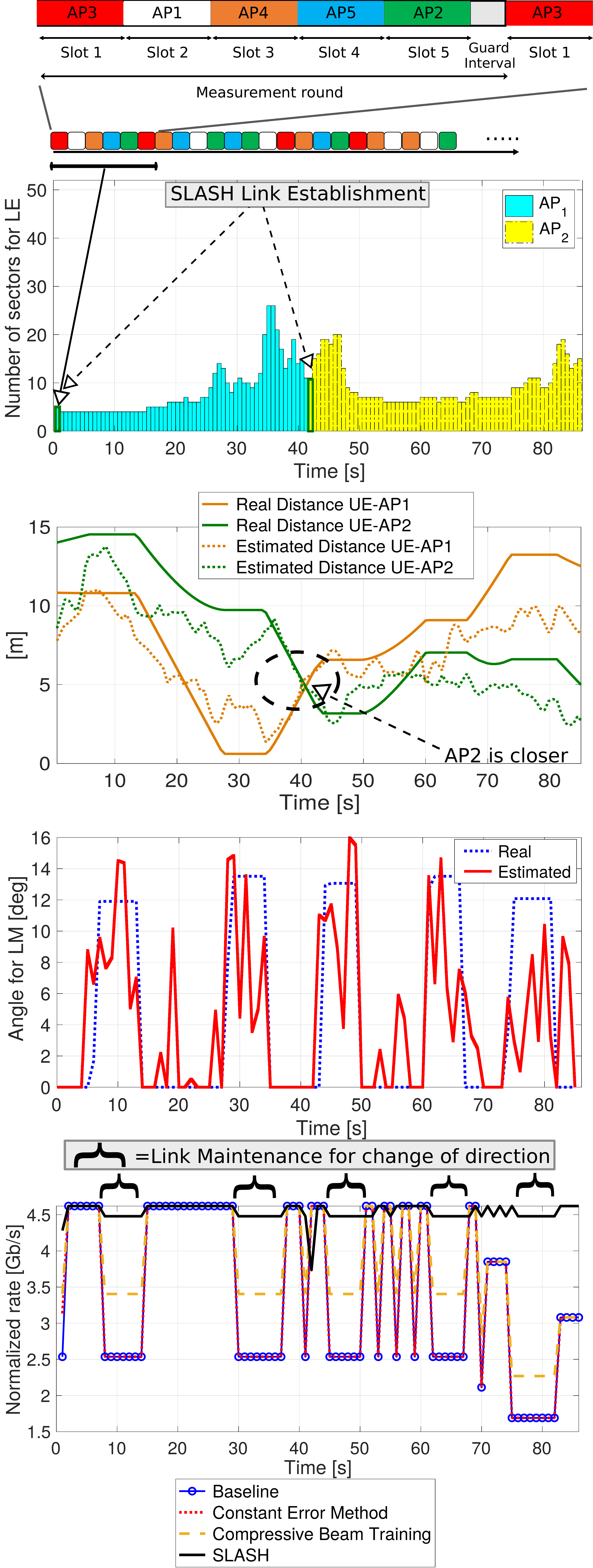}
    \vspace{-2mm}
 \caption{Context-aware information (number of sectors for LM, angle 
error for LE, and distance estimate) system, and data rate of SLASH and other algorithms.}
 \vspace{-5mm}
 \label{fig:context-awareness}
\end{figure}

\subsection{SLASH for a mobile user}

%
%
We study the performance of SLASH with a user moving along the 
mobility pattern in~Fig.~\ref{fig:scenario}, which includes straight-line paths and rotations. 
For the experimental trace, the mobile user holds the 
UE, walks along the trajectory at an approximated speed of $0.5$\,m/s and rotates
at an approximated rotation speed of $0.35$\,rad/s.
In the map in Fig.~\ref{fig:scenario}, we show the real trajectory of the user with the 
blue dashed line, while the one 
estimated by our positioning system is shown as black dotted line. Furthermore we use an orange 
diamond for the position where the user performs a handover to a second AP when the estimated ToF range
indicates that it is closer to the user. 
For this study, we consider two APs ($\mathrm{AP}_1$ and $\mathrm{AP}_2$) for the user to switch between, as the other three 
APs are in office 
rooms and they could not provide good coverage along the trajectory at mmWave frequency. 

On the top of Fig.~\ref{fig:context-awareness}, we show the time division scheduler of the positioning system. 
Estimating 
the distance from each AP, the estimated position is then 
available. The median error of the positioning system along the trajectory is $2.3$\,m. 
Positioning data is processed for the 
calculation of the angle error $\theta_p$. The latter is used to determine the number 
of sectors for link establishment, based on the 7$^\circ$ antenna beamwidth.
This results in the temporal evolution of the number 
of sectors shown in the second plot from the top in Fig.~\ref{fig:context-awareness}. Referring to Fig.~\ref{fig:scenario}, the 
UE is at time zero in the position UE1 marked with a green cross. Here it requests the 
access to the mm-wave network and the link 
establishment procedure of SLASH is triggered in order to establish a 
reliable AP-UE link.  For completeness we show 
how many sectors would be probed over the entire trace, if the link establishment procedure would be executed at that time.
 After approximately $41$\,s, a link establishment procedure is 
triggered as  $AP_2$ has better signal quality. This is performed as soon as the mm-wave link quality
decreases using ToF distance estimates (third figure from the top in 
Fig.~\ref{fig:context-awareness}).

\balance

Using the TOF range 
measurements as input for the rotation estimator, we provide the angle for link maintenance. TOF range measurements are collected 
in each AP over the last second, and rotation estimates from different APs are integrated over the observation period. The 
resulting estimated rotation over time is shown in the fourth figure from the top in 
Fig.~\ref{fig:context-awareness}, where we have a median error of 1.54$^\circ$ along the whole trajectory.

We finally study the normalized data rate along the whole trajectory for four 
different algorithms. We recall that we resort to simulations using a trace-driven approach since our mechanical 
rotation does not allow our system to operate in real-time. The results are
shown in the plot at the bottom in 
Fig.~\ref{fig:context-awareness}. SLASH tries to maintain the highest data rate,
and the algorithm is called as soon as 
the RSS drops below the threshold for communication at the highest data rate.
In addition, the user performs the handover to $AP_2$ 
according to the distance estimates of SLASH. Other strategies performs a handover as soon as the mm-wave link
is below the threshold for communication. However, this does not occur in the open space under study. The plot shows the time when 
link maintenance is 
triggered due to the rotation. The ''Baseline``, 
''Constant Error Method`` and 
``Compressive Beam Training'' approaches are not able to detect rotation events and, therefore, 
need to resort to the same full beam search used for link establishment.  During link maintenance, SLASH 
achieves a data rate 67\% higher than ``Compressive Beam Training'' 
and 121\% higher than other strategies.

\section{Related work}\label{sec:related_work}

The problem of fast mm-wave link establishment and maintenance is widely discussed in the literature. 
A comparative analysis of initial access techniques in mm-wave networks is presented \mbox{in \cite{mezzav-1}}.
Simultaneous transmissions from multiple direction-coded beams to accelerate the beam search are exploited 
in~\cite{beam-training,5-abf}. 
In \cite{katabi_beampatterns} the authors use multi-lobe antenna patterns with 
random phase shifts for the beam training, which enables compressive sensing approaches to determine AoA and AoD. 
This approach requires arbitrary phase shifters over the antenna elements, which is not supported by standard 
off-the-shelf mm-wave devices that operate with how-resolution (2-bit) RF phase shifters~\cite{6847111}. 

In~\cite{sur201560}, a link-level measurement study of indoor 60\,GHz networks using a software-radio platform revealed several 
challenges related to human blockage and device motion and how they affect the design of MAC protocols. Some of the findings of 
that work have been applied to extract context information from mm-wave networks in BeamSpy~\cite{Sur:2016:beamspy}. BeamSpy 
presents a beam tracking prediction mechanism based on the channel sparsity and spatial correlation of the 60 GHz link.
The proposed approach works only in static conditions, while we target mobile users.
Mm-wave and sub-6 GHz WiFi WiFi context information can complement each other and can enrich the set of inputs provided for the 
design of MAC protocols. \cite{Steinmetzer2017} uses compressive sensing to sweep only through a subset of probing sectors.
However, the authors need still to probe a random subset of beams, while SLASH exploits knowledge from sub-6 GHz WiFi WiFi context 
information to probe in the direction that is statistically more likely to contain the beam with the highest RSS.


Only very few works in the literature address the problem of fast beam search in realistic, dynamic scenarios with node mobility. 
In~\cite{haider2016mobility}, a protocol for mobility resilience and overhead constrained adaptation for directional 60\,GHz 
links 
is presented. A ``beam sounding'' mechanism is introduced to estimate the link quality for selected beams, and identify and adapt 
to link impairments. In~\cite{loch2017zero} the authors develop a zero overhead beam tracking mechanism 
that uses two different beam patterns during the preamble of the frames. This allows to estimate user rotation and movement. 
The approach does not use context information from out-of-band channels, and it requires changes in the preamble structure.
 In~\cite{nitsche2015steering} 
an algorithm that removes in-band overhead  for directional mm-wave link  establishment is proposed. However, the study focuses 
only on static scenarios without rotation, and requires at least five antennas in both the sub-6 GHz transmitter and receiver
 to achieve a similar angle error as in our system.

\cite{patra2016pimrc} relies on 
gyroscope sensors to reduce the beam search under mobility. In this paper, we instead present a method that uses sub-6 GHz ToF 
ranges
for mm-wave link maintenance. The coupling of legacy technology together with mm-wave 
frequencies in order to improve the beam search has been studied in some recent work. In~\cite{zorzi_eucnc}, 
it is shown that beam training and cell discovery, respectively, can be accelerated
assuming 
the availability of Global Positioning System (GPS) information about device locations. \cite{zorzi_eucnc} treats the location 
accuracy as a given parameter without 
developing any angle error model based on the desired confidence level of the estimated position. Our WiFi ToF positioning system 
can be complemented with 
solutions that perform channel switching to achieve a finer-grained timing 
information for ranging~\cite{xiong2015tonetrack} and with 
solutions that efficiently select the best communication technology between low frequency and mm-wave~\cite{zhang2017}.

\section{Conclusion}\label{sec:conclusion}

In this work we investigated how context information extracted from a  sub-6 GHz WiFi ToF radio ranging and positioning 
system can help 
to speed-up the beam training process in mm-wave networks. Our system did not use any inertial sensors but just radio 
signals. Context is extracted from a method to infer the speed of rotation of the device using 
ranging measurements, and 
from  a
closed-form expression of the statistical angle error model based on position and position error estimates. We then 
introduced SLASH, to perform statistical beam search for link establishment and maintenance, exploiting
 the ties between the quasi-reciprocity of the mm-wave channel and 
the user's position to further speed up the link establishment, and the ToF distance estimates to probe the presence of APs with 
better quality.
We have shown through extensive experiments with a multi-band system that SLASH can significantly increase the data rate 
for static and mobile users compared to prior work. Our mechanism to select 
the number of sectors to scan based on context information can 
directly be integrated in 
multi-band WiFi devices.

\appendix
\section{Appendix} \label{sec_appendix}

For the formal definition of HDOP, we denote 
$$\bold{P}^\text{AP} = \begin{bmatrix} \bold{p}_1^\text{AP} 
& \bold{p}_2^\text{AP} & \hdots & 
\bold{p}_N^\text{AP}\end{bmatrix} \in \mathbb{R}^{3\times N}$$ the matrix containing the AP coordinates. We can then define the 
matrix $A$ containing the unit vectors of the 
direction between each AP and the UE:
\begin{equation}
\bold{A} = 
 \begin{bmatrix}
  (\bold{p}_1^\text{AP} - \hat{\bold{p}}^\text{UE})/ {\lVert \mathbf{\bold{p}_1^\text{AP}-\hat{\bold{p}}^\text{UE}} \rVert} & -1 
\\
  (\bold{p}_2^\text{AP} - \hat{\bold{p}}^\text{UE})/ {\lVert \mathbf{\bold{p}_2^\text{AP}-\hat{\bold{p}}^\text{UE}} \rVert} & -1 
\\
  \vdots  &  \vdots  \\
  (\bold{p}_N^\text{AP} - \hat{\bold{p}}^\text{UE})/ {\lVert \mathbf{\bold{p}_N^\text{AP}-\hat{\bold{p}}^\text{UE}} \rVert} & -1 
 \end{bmatrix}
 \label{eq:hdop_formulaA}
\end{equation}
and formulate the matrix $\bold{Q} = (\bold{A}^T\bold{A})^{-1}\in \mathbb{R}^{4\times4}$. The HDOP can be then computed as:
\begin{equation}
\mathrm{HDOP} = \sqrt{\bold{Q}_{11}+\bold{Q}_{22}}\, .
\label{eq:hdop_formula}
\end{equation}

\bibliographystyle{IEEEtran}
\bibliography{IEEEfull,mybib}

\end{document}